\documentclass[10pt, journal, a4paper, final, nofonttune]{IEEEtran}
\usepackage{amssymb,amsmath,amsthm,epsfig,graphics,float,subfigure, mhchem, siunitx}

\begin{document}
\title{The Biosensor based on electrochemical dynamics of fermentation in yeast \emph{Saccharomyces Cerevisiae}}
\author{Serge Kernbach$^1$, Olga Kernbach$^1$, Igor Kuksin$^1$, Andreas Kernbach$^1$, Yury Nepomnyashchiy$^1$, Timo Dochow$^2$, Andrew Bobrov$^3$\\[3mm]
\small $^1$CYBRES GmbH, Research Center of Advanced Robotics and Environmental Science, Melunerstr. 40, 70569 Stuttgart, Germany\\
\small $^2$IFFP GmbH, Oberaustrasse 6b, D-83026 Rosenheim, Germany, $^3$Orel State University, Komsomolskaya, 95, 302026 Orel, Russia
\thanks{$^1$Corresponding author: serge.kernbach@cybertronica.de.com}
\vspace{-5mm}}

\date{}
\maketitle
\thispagestyle{empty}

\begin{abstract}
The zymase activity of the yeast \emph{Saccharomyces cerevisiae} is sensitive to environmental parameters and is therefore used as a microbiological sensor for water quality assessment, ecotoxicological characterization or environmental monitoring. Comparing to bacterial bioluminescence approach, this method has no toxicity, excludes usage of genetically modified microorganisms, and enables low-cost express analysis. This work focuses on measuring the yeast fermentation dynamics based on multichannel pressure sensing and electrochemical impedance spectroscopy (EIS). Measurement results are compared with each other in terms of accuracy, reproducibility and ease of use in the field conditions. It has been shown that EIS provides more information about ionic dynamics of metabolic processes and requires less complex measurements. The conducted experiments demonstrated the sensitivity of this approach for assessing  biophotonic phenomena, non-chemical water treatments and impact of environmental stressors. 
\end{abstract}

\section{Introduction}

Yeasts are single-celled microorganisms from the fungus kingdom. The yeast species \emph{Saccharomyces cerevisiae} (S.cer.), known as Baker's yeast, is widely used in winemaking, baking, and brewing  production, as well as in scientific research -- the genome was completely sequenced in 1996 \cite{Goffeau25101996}. The ethanol fermentation of S.cer. converts sugar in forms of glucose, fructose, and sucrose into ATP (Adenosine triphosphate) with ethanol and carbon dioxide as side-products. Beside glucose (fructose, sucrose) fermentation, S.cer. can perform fermentation of maltose -- both are denoted as zymase and maltase activities of yeasts \cite{JIB:JIB3223}, \cite{JIB:JIB2843}.

Tests on zymase activity are defined as a rate of gas production by yeasts working in sugar solutions at fixed temperature \cite{JIB:JIB2789}. Zymase activity, beside sugar concentration and temperature, depends on other parameters. For instance, such tests performed in production environments \cite{Kornienko_S04}, \cite{Kornienko_S03A} enable estimating the quality of yeasts, parameters of metabolic processes; in combination with optical measurements (e.g. flocculation by turbidity sensors) -- detecting parameters of technological processes in baking and brewing production \cite{Herrera91}, \cite{Stratford92}. Several available on the market devices \cite{ANKOM17}, \cite{ABER17} are developed and used for these purposes.

Outside of food and beverage production, tests based on zymase activity are also used in detecting different environmental stimuli, estimating the quality of water \cite{Bobrov02en}, \cite{Bobrov01en} or detecting non-chemical water treatments \cite{Harfst10}, \cite{PIYADASA201719}, \cite{doi:10.1021/ja972171i}. Tasks of ecotoxicological characterization \cite{PANDARD2006114} and detection of weak environmental stressors (WES) are similar, however differ in their purpose: the first one estimates the biological toxicity, whereas the second one measures the impact of specific non-toxic stimuli. Examples of WES are weak EM emissions, pulsed electric and magnetic fields \cite{ALIMI20091327}, distant biophotonic signalling \cite{prasad2004introduction}, light, infra-/ultra- sound, or experimental factors related to magnetic vector potential \cite{Rampl2009}, \cite{Puthoff98} and the Aharonov-Bohm effect \cite{Aharonov59}. In general, application of microorganisms, and especially S.cer. \cite{s8106433}, as biological sensors represents the well-known approach in different environmental and technological branches \cite{campbell2012sensor}, \cite{KARUBE198869}, \cite{doi:10.1021/jf9911794}. The advantage of such sensors consists in a high sensitivity, capability to sense a wide range of stimuli that impact the metabolism of microorganisms, low costs and suitability for express analysis. In previous publications we demonstrated that the ethanol fermentation of S.cer. depends on exposure by EM fields and laser radiation \cite{Kernbach14Biominimalen}. The prototypes of corresponding devices and measurement methodologies are developed and tested in multiple experiments.

This paper extends these works. The first developed system has 16 integrated high-resolution pressure sensors, intended for simultaneous measurement of 8 control and 8 experimental populations, see Fig.\ref{fig:Devices}(a). Since the production of $CO_2$ and $C_2 H_5 OH$ by yeast changes the ionic content of water, we tested the second measurement system based on electrochemical impedance spectroscopy (EIS) \cite{Kernbach17water}, \cite{Kernbach17EISen}. The application of EIS in fermentation control is known \cite{s150922941}, \cite{JEB664}, the approach shown here extends the state of the art towards high-resolution differential measurements. In this work we developed and tested two thermostablized EIS systems for biological measurements: with small \SI{15}{\milli\litre} containers used in the water research, see Fig.\ref{fig:Devices}(b), and with large \SI{100}{\milli\litre} containers, see Fig.\ref{fig:Devices}(c). The performed experiments consist of three groups: control attempts with equal conditions for both populations, different water treatments, and exposure of yeast by WES, produced, in particular, by technologies available on the market. The main goal of this work is to demonstrate the applicability of biosensing approach for environmental measurements, to show the advantages and disadvantages of pressure and EIS techniques and to estimate their sensitivity and accuracy.

\section{Sensing weak environmental stressors by microorganisms -- state of the art}
\label{sec:fermentation}

One of the most famous approaches is the Chizhevsky-Velhover effect \cite{Chizhevsky36} -- the cytochemical metachromatic reaction of polyphosphate-containing volutin granules of microorganisms. This effect appears as cyclical metachromatic staining (change of colors) related to cosmic emissions (probably in mm area \cite{doi:10.1117/12.2017989}), discovered in 30x of XX century. Multiple replications in 60x-70x \cite{Gorshkov72} and in 2001-2015 \cite{Gromozova15} confirmed this effect. Experiments described in  \cite{Gromozova15} have been performed with the yeasts \emph{Saccharomyces cerevisiae}. Currently, the heliobiology (study of the Sun effects, geomagnetic activity and EM fields on biology) is well-accepted scientific branch with a large number of publications \cite{konig2012biologic}, \cite{Palmer2006}.

Another well known approach of environmental sensing with microorganisms is based on bacterial bioluminescence \cite{Nealson79}, in particular of bacteria \emph{E. coli} \cite{Ritchie01062003}. For instance, this approach was utilized for measuring biological effects of magnetic vector potential  \cite{Anosov03en}, \cite{Rampl2009}. The work \cite{NUNES-HALLDORSON2003} describes different applications of bioluminescent bacteria as environmental biosensors, it needs to note such bacteria exist in parasitic forms and as opportunistic pathogens. There are multiple commercial test systems based on gene-modified \emph{E. coli} (e.g. 'Ecolum') \cite{Deryabin04}.

Microorganisms are used in the research of weak biophotonic emissions \cite{Gurvich44en}, \cite{Popp94}. For instance, in the work \cite{Kaznacheev81en} different tissues (fibroblasts, muscle cells, kidney epithelium) are investigated; the systematic review of distant biophotonic WES can be found in \cite{Ives14}, \cite{Hunt09}. Biophotonic emission of S.cer. was investigated multiple times, e.g \cite{MASLANKA201855}, \cite{Musumeci99}. Ecotoxicological characterization of e.g. wastes is performed by such tests as growth inhibition, mobility inhibition or the rate of lethality in populations of different microorganisms \cite{PANDARD2006114}, \cite{ROJICKOVAPADRTOVA1998495}. The toxicity detection is specific for each type of contamination, thus tests select the most appropriate species (or set) of microorganisms, for instance the mobility tests of \emph{Spirostomum amobiquum} was utilized for detection of EM stimuli \cite{Anosov03en}.

\begin{figure}[htp]
\centering
\subfigure[]{\includegraphics[width=0.4\textwidth]{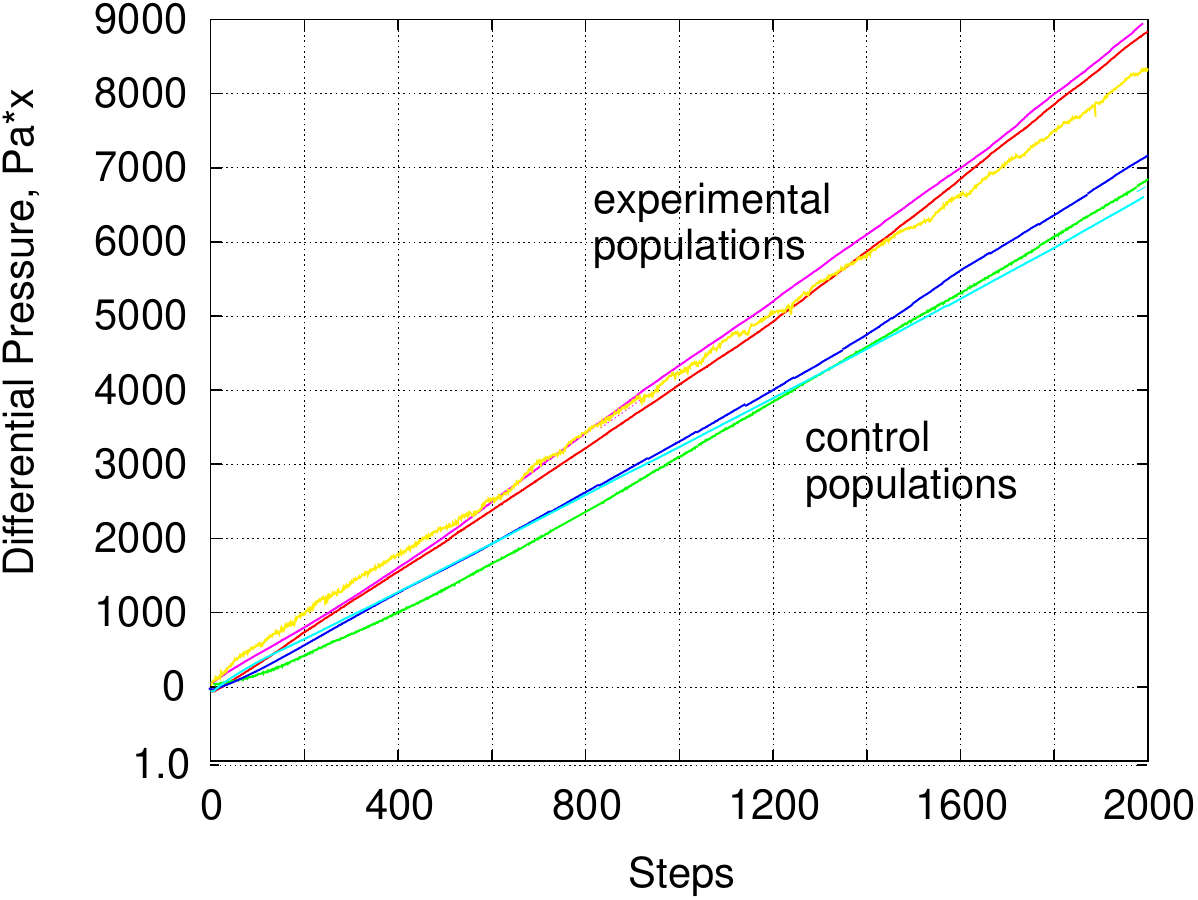}}
\subfigure[]{\includegraphics[width=0.4\textwidth]{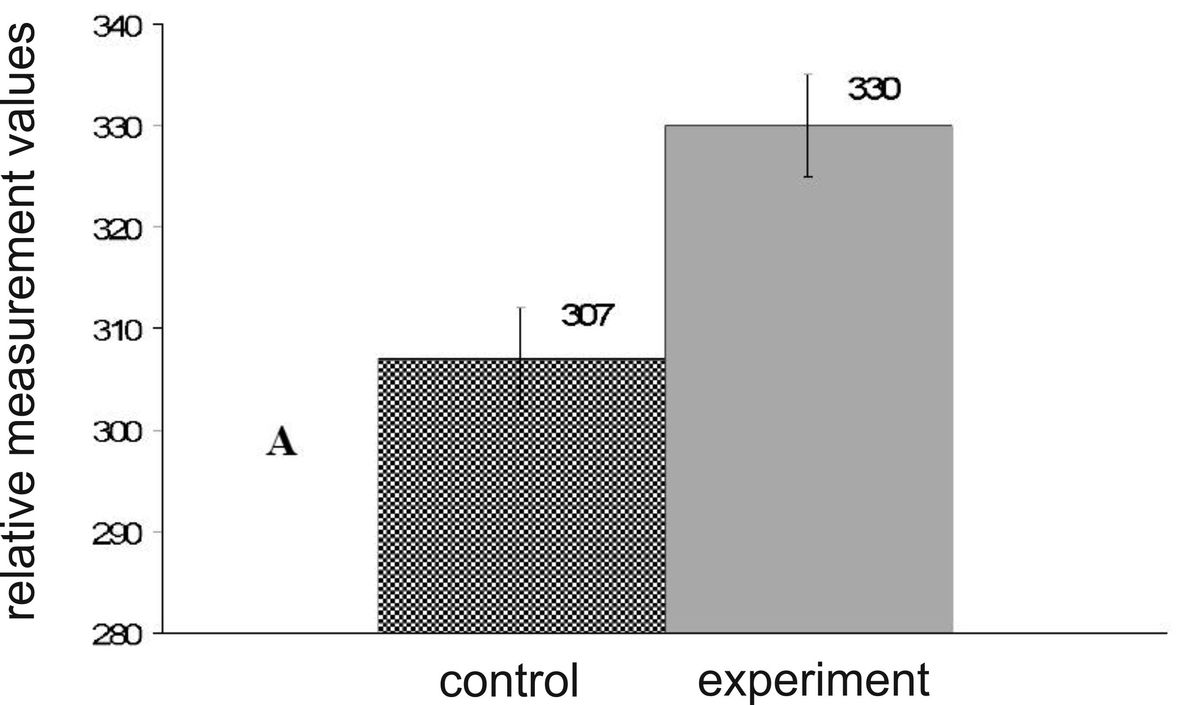}}
\caption{\small Examples of measurements from previous versions of pressure sensing system, yeast are exposed by WES \cite{Kernbach12JSE}, the difference in pressure of experimental and control populations is well visible: \textbf{(a)} (Stuttgart) 6-channels pressure sensing system based on the Murata SCP1000-D11 sensors, images from \cite{Kernbach14Biominimalen}; \textbf{(b)} (Orel) 15-channels pressure sensing system, image from \cite{Bobrov01en}. \label{fig:resStuttgart}}
\end{figure}

The environmental sensing with microorganisms can be conducted with MFC (microbial fuel cell) technology. The first MFC appeared in 70x (the operating principles are demonstrated in the early 20th century). Despite it is developed as the energy producing method, the electrical parameters of MFC reflect the metabolic processes in microorganisms and thus can be used for sensing purposes \cite{bath50782}, e.g. as BOD (biological oxygen demand) sensors \cite{Kim2003} or for monitoring the water quality \cite{bios5030450}.

Tests with zymase activity of \emph{Saccharomyces cerevisiae} and development of corresponding me\-tho\-dologies and devices started first by research teams in Orel (Russia) \cite{Bobrov02en}, \cite{Bobrov01en} and then in Stuttgart (Germany) \cite{Kernbach14Biominimalen}, \cite{Kernbach15en} with pressure sensing approach. Goal of these measurements consisted in estimating the biological impact of WES, produced by different experimental devices, as described e.g. in \cite{Kernbach12JSE}. Yeasts were exposed in closed single- and double-wall metal containers having a passive thermostabilization. Examples of such measurements are shown in Figs. \ref{fig:resStuttgart}. These experiments replicated and confirmed results obtained in both laboratories and represent the background works for this paper.

\section{Device description and methodology}
\label{sec:device_description}

\subsection{Measurement equipment}

\begin{figure}[htp]
\centering
\subfigure[]{\includegraphics[width=.245\textwidth]{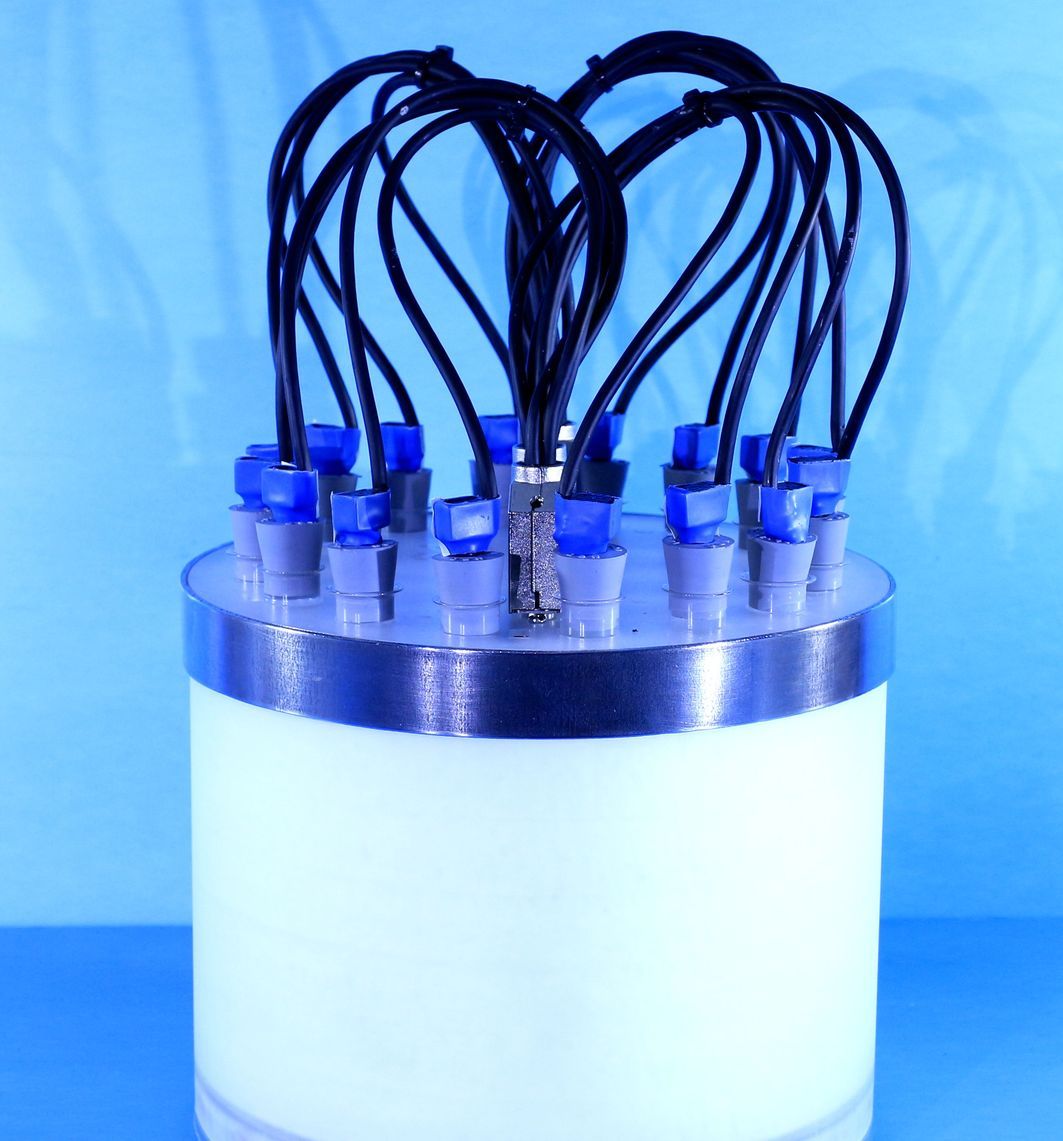}}~
\subfigure[]{\includegraphics[width=.245\textwidth]{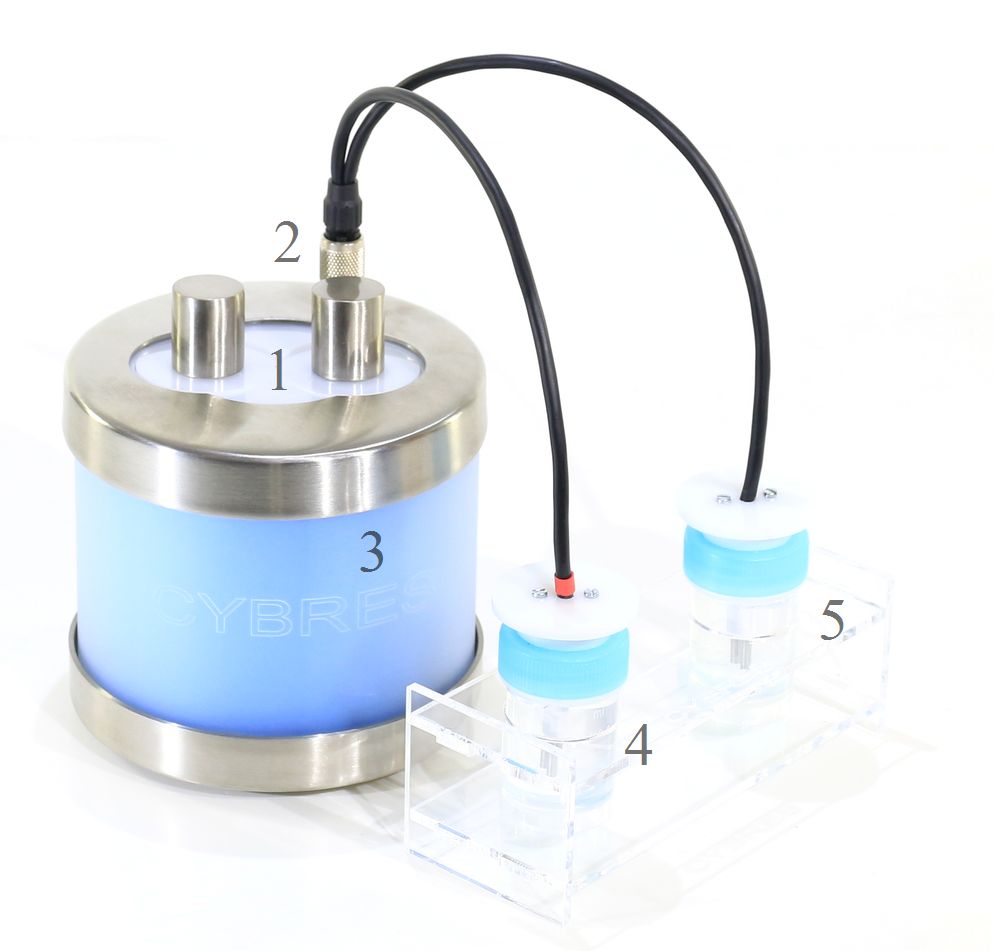}}
\subfigure[]{\includegraphics[width=.245\textwidth]{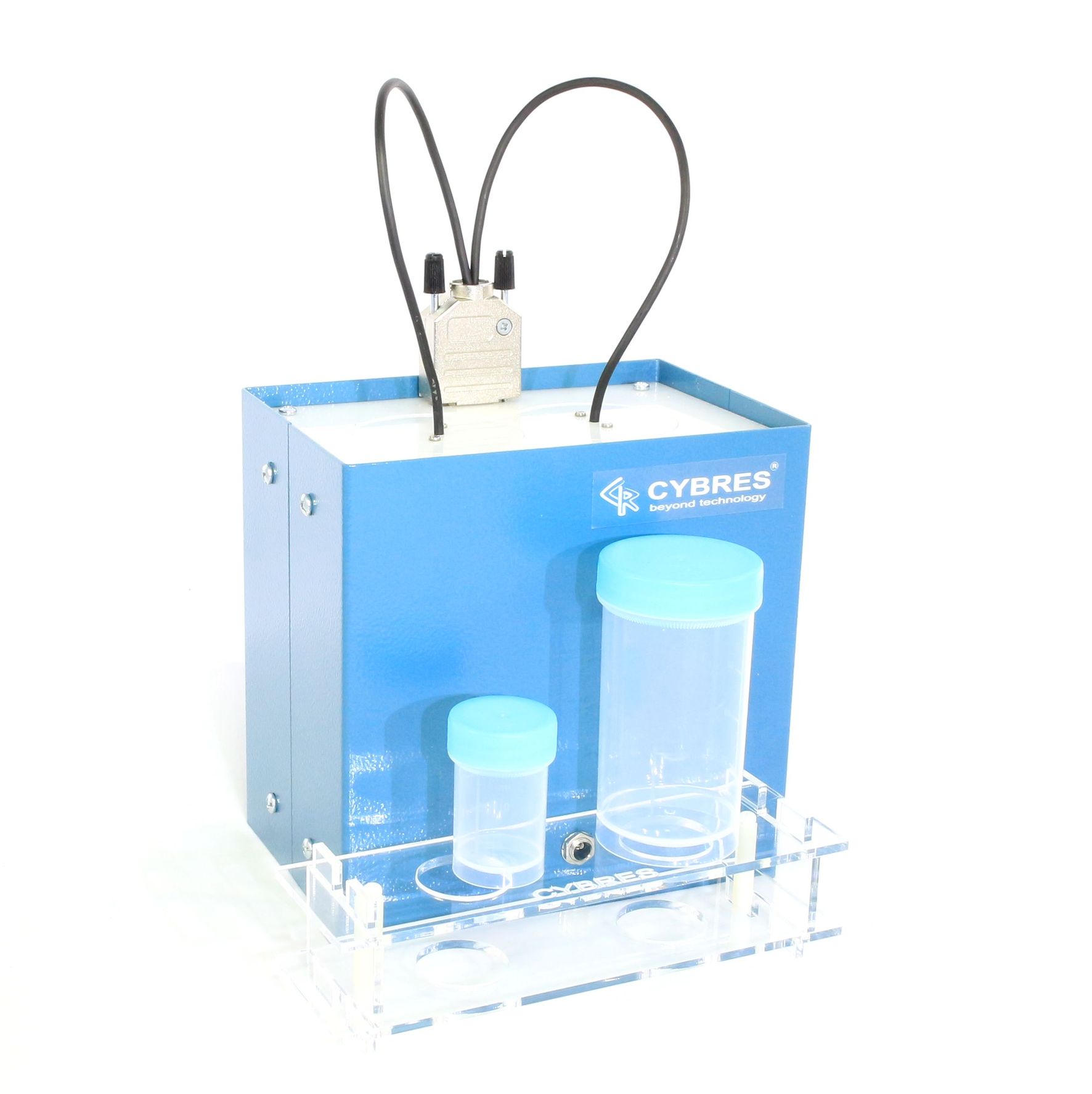}}~
\subfigure[]{\includegraphics[width=.245\textwidth]{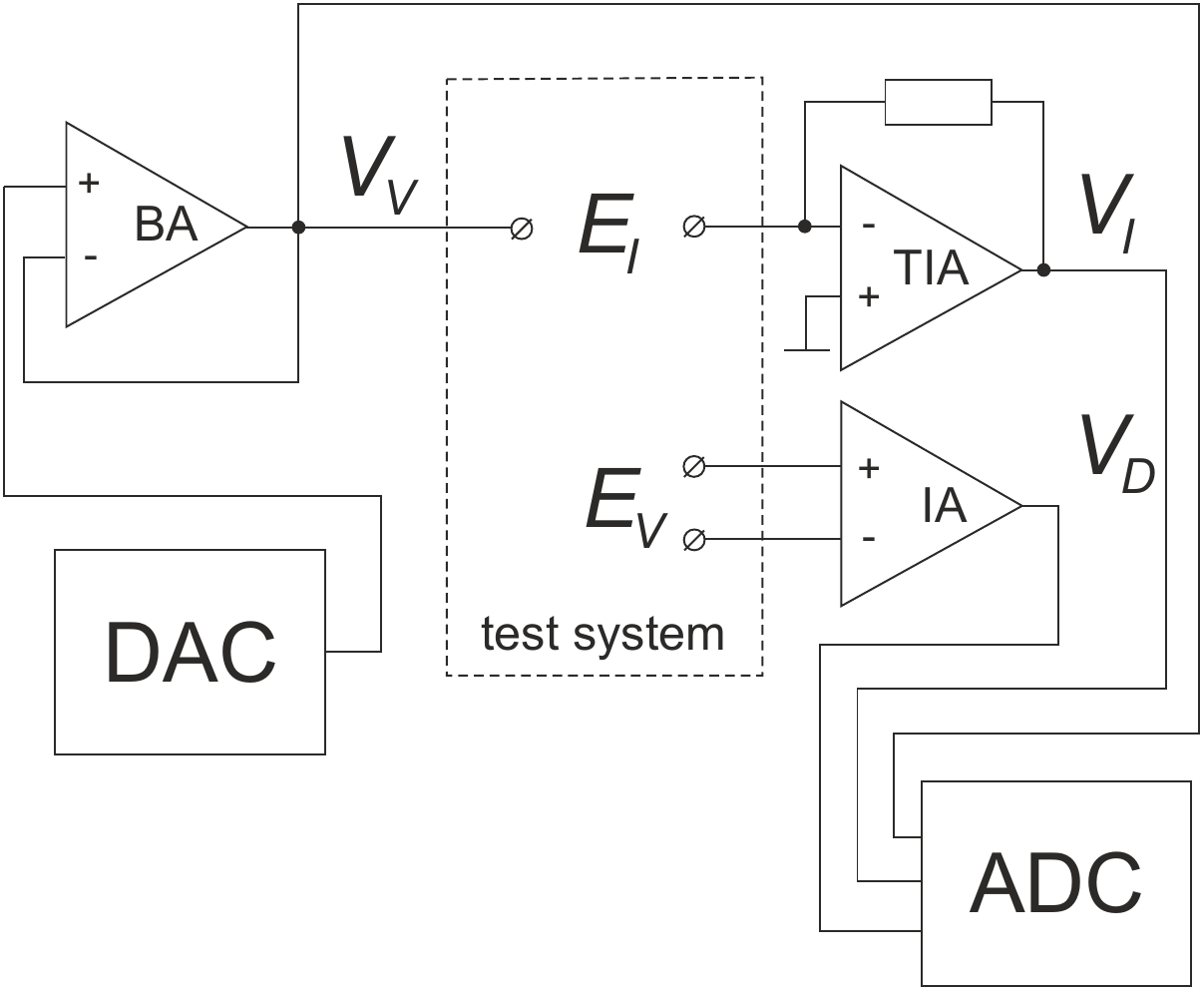}}
\caption{\small \textbf{(a)} The 16-channel pressure sensing system; \textbf{(b)} The differential EIS spectrometer with temperature stabilization of samples and electronic components; \textbf{(c)} Adapted version of the EIS device with \SI{100}{\milli\litre} containers for the biosensor purposes; \textbf{(d)} Structure of one EIS measurement channel, see description in text. \label{fig:Devices}}
\end{figure}

\emph{Pressure measurement device.} The device represents a 16-channel measuring system with Honeywell pressure sensors, see Fig. \ref{fig:Devices}(a). Since the fermentation depends on temperature, the reagent chambers have a built-in thermostat, which maintains a constant temperature. Alcohol fermentation is carried out in conventional test tubes, which are closed by threaded caps with built-in Honeywell 26PC gauge pressure sensors. They are factory calibrated and temperature compensated. Each sensor contains four active piezoresistors that form a Wheatstone bridge and the output signal is given with \SI{0.25}{\%} accuracy \cite{Honeywell2015}. The system uses 8 control and 8 experimental populations. The device is connected to the computer via USB interface, one measurement lasts about \SIrange[range-phrase = --]{60}{90}{\minute}. The absolute maximum value of the measuring pressure is \SI{200}{\kilo\pascal}. The sensitivity threshold of the device is \SI{3.15}{\micro\volt} or \SI{193}{\milli\pascal}. The measured noise level (P-P) is \SI{+-1}{\pascal}.
The chamber and PCB temperatures are stabilized by PID controller in linear mode to prevent the noise generation, the temperature instability is \SI{+-0.005}{\degreeCelsius}. All mechanical impacts on the device are monitored by accelerometer and magnetometer.

\emph{EIS measurement device}, see Fig. \ref{fig:Devices}(b), is already described in e.g. \cite{Kernbach17EISen}, \cite{Kernbach17water}. This is a compact device for differential EIS with two and four electrodes measurement with thermal stabilization of the electronic system and samples. The EIS device shown in Fig. \ref{fig:Devices}(c) represents results described in this work; it uses larger containers and adapted thermostabilization, see more in Sec. \ref{sec:experiments}. Both EIS meters use an auto-balancing bridge, where a test system is excited by the voltage $V_V$, see Fig. \ref{fig:Devices}(d). The signal waveform for $V_V$ is generated by DAC and is buffered by the buffer amplifier (BA). The flowing current $I$ through the test system is converted into a voltage $V_I$ by the transimpedance amplifier (TIA). Synthesis of the signal $V_V$ occurs by Direct Digital Synthesis (DDS) with 32-bit frequency resolution, the signals are digitized by two synchronous \SI{1.2}{MSPS} ADCs for simultaneous sampling of $V_V$ and $V_I$ signals. The EIS meter uses an external analog circuitry for impedance matching. The electrode pair $E_I$ is used for the current sensing $V_V \rightarrow V_I$ (two electrode system). Another electrode pair $E_V$ is used to sense a differential potential with the instrumental amplifier (IA, four electrode system). Additionally to EIS data, the system records parameters from 14 additional sensors (e.g. temperature, light, air humidity/pressure, accelerometer, magnetometer) with real time stamps. Both channels are equipped with the resistive temperature sensors (\SI{+-1}{\%} accuracy) for monitoring the temperature of fluids in biosensor applications.
\begin{figure}[h!]
\centering
\subfigure[]{\includegraphics[width=.49\textwidth]{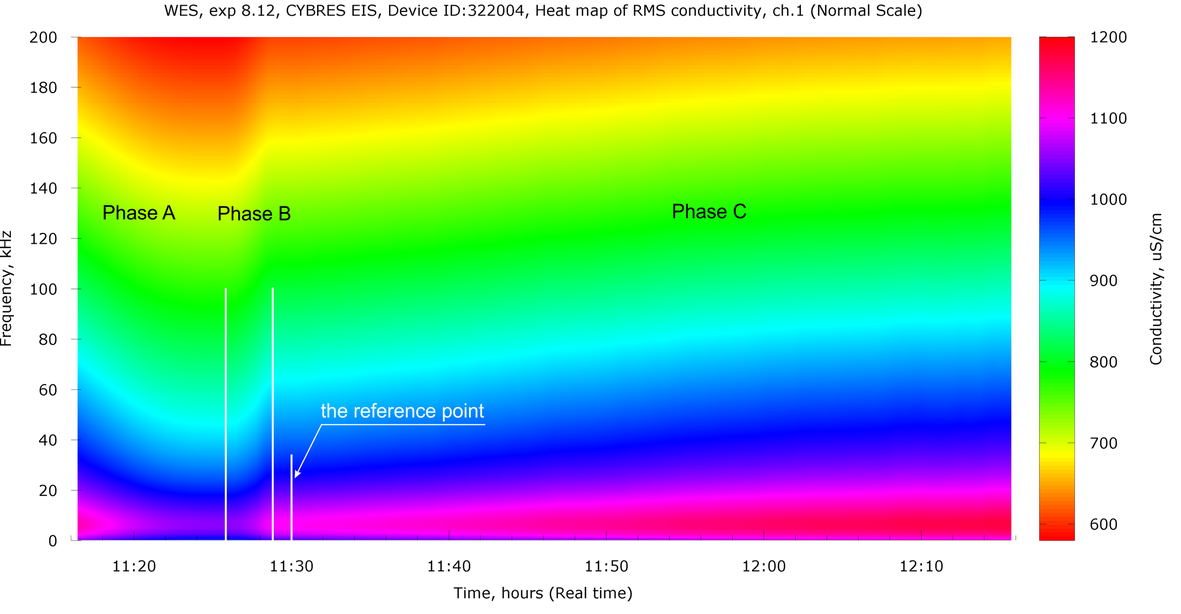}}
\subfigure[]{\includegraphics[width=.49\textwidth]{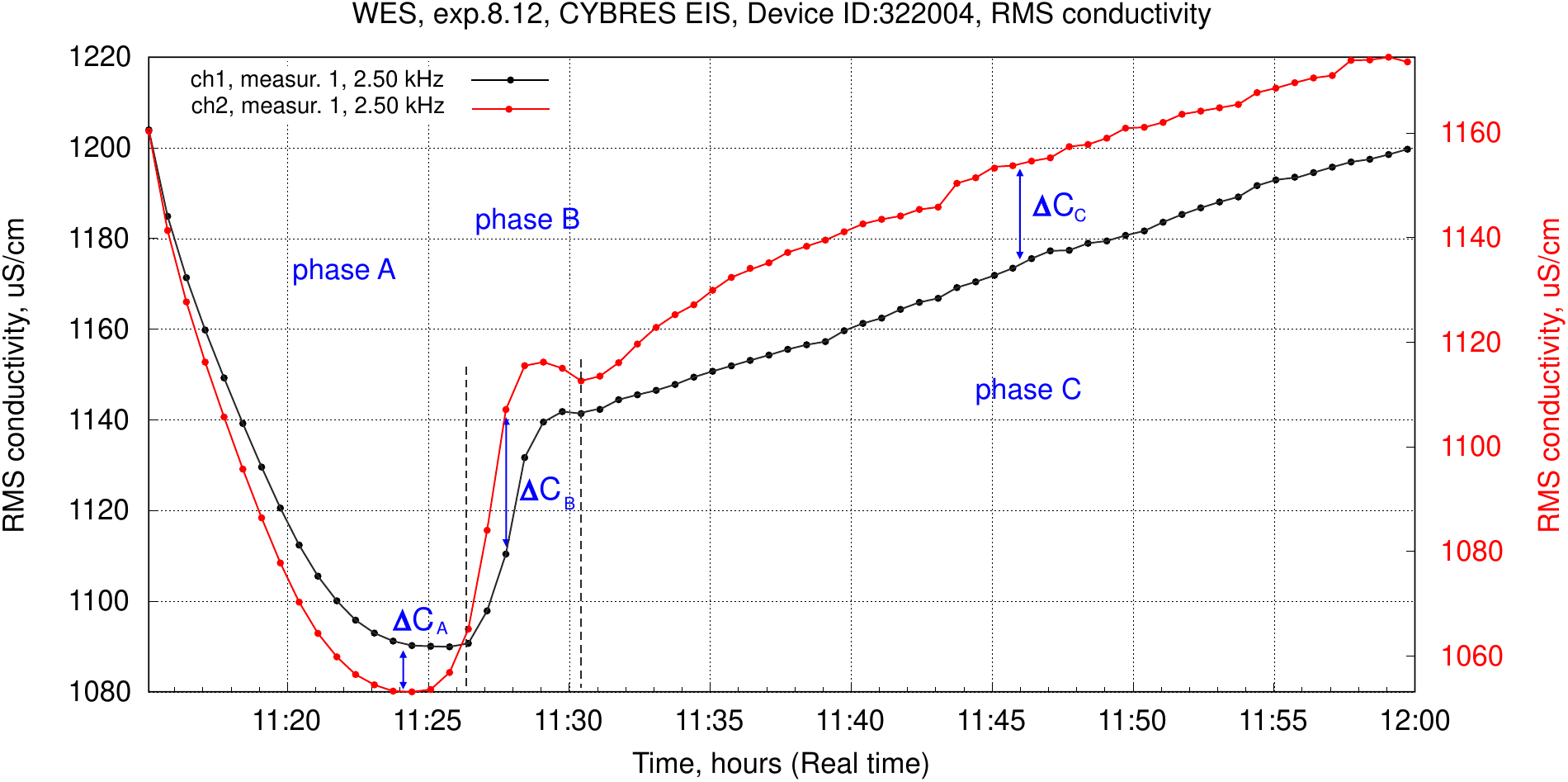}}
\caption{\small Electrochemical phases of fermentation dynamics: \textbf{(a)} EIS data (3D multi-frequency representation) from the channel 1, the phase A -- sedimentation process, the phase B -- fast dynamics, the phase C -- slow dynamics; \textbf{(b)} 2D plot at \SI{2.5}{\kilo\hertz} of the channels 1 and 2 from (a), the channel 2 is impacted by WES, different EIS dynamics of impacted and non-impacted channels is well visible in all phases of fermentation process.
\label{fig:EIS}}
\end{figure}

\subsection{Electrochemical dynamics of \textit{Saccharomyces cerevisiae}}

Spectrograms from one typical EIS measurement (the channel 2 is impacted by WES, the channel 1 is a control channel) are shown in Fig. \ref{fig:EIS}(a). The EIS behavior has three different regions -- the phases A, B, and C. To understand this dynamics, we note that yeasts perform aerobic fermentation with the dissolved $O_2$
\begin{equation}
\label{eq:1}
C_6 H_{12} O_6 + 6 O_2 \rightarrow 6 CO_2 + 6 H_2 O,
\end{equation}
which is energetically more preferable. In absence of oxygen, yeasts switch to anaerobic fermentation
\begin{equation}
\label{eq:2}
C_6 H_{12} O_6 \rightarrow 2 CO_2 + 2 C_2 H_5 OH,
\end{equation}
which includes different pathways to synthesize necessary metabolic products \cite{dickinson2004metabolism}. Both types of fermentation produce $CO_2$, which dissolves to  $H^+$ and $HCO_3^-$ ions
\begin{equation}
\label{eq:3}
CO_2 + H_2 O \rightarrow (H_2 CO_3 ) \rightarrow H^+ + HCO_3^-.
\end{equation}
\begin{figure}[t!]
\centering
\subfigure[\label{fig:EIS2a}]{\includegraphics[width=.49\textwidth]{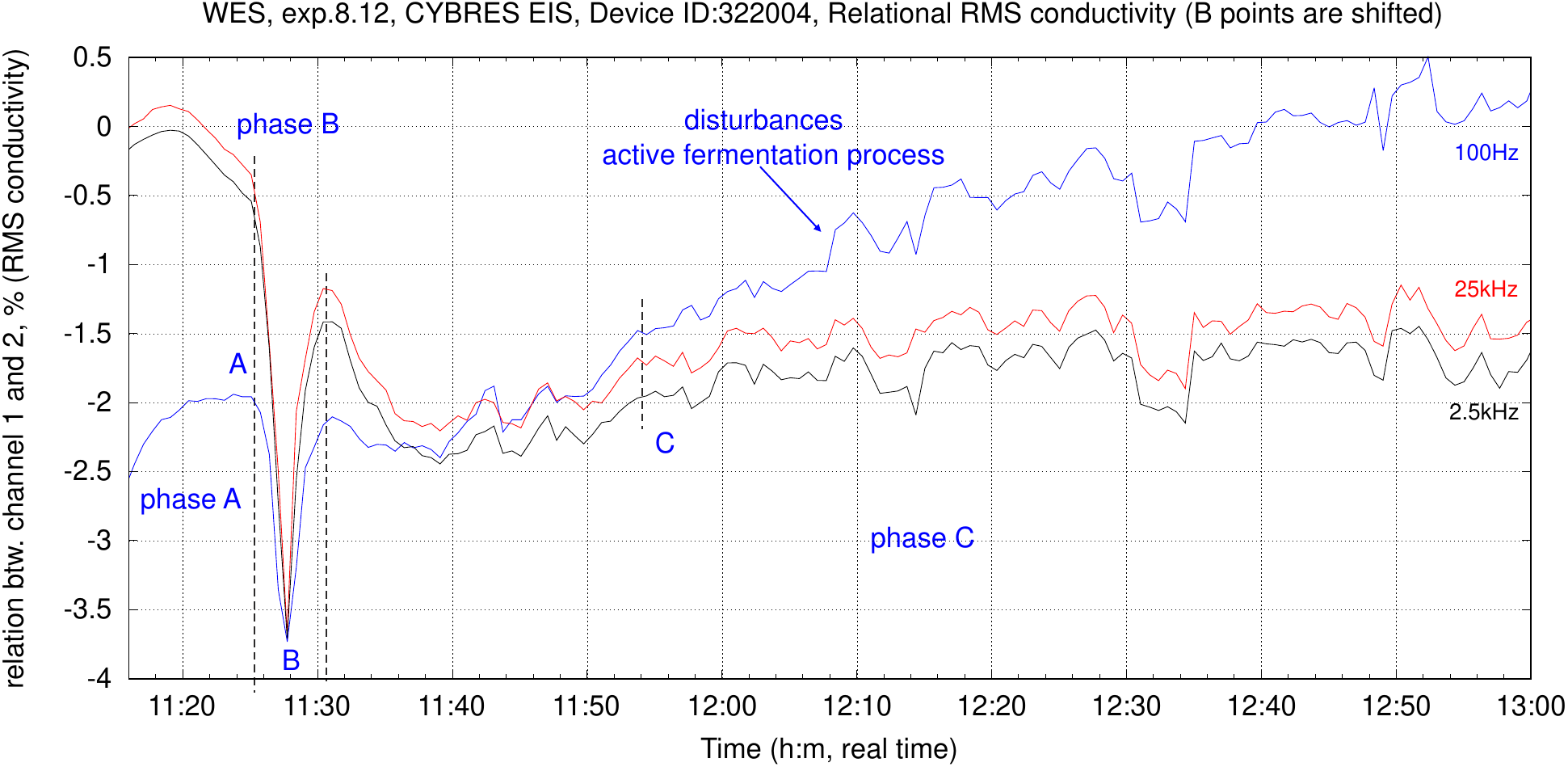}}
\subfigure[\label{fig:sediment}]{\includegraphics[width=.4\textwidth]{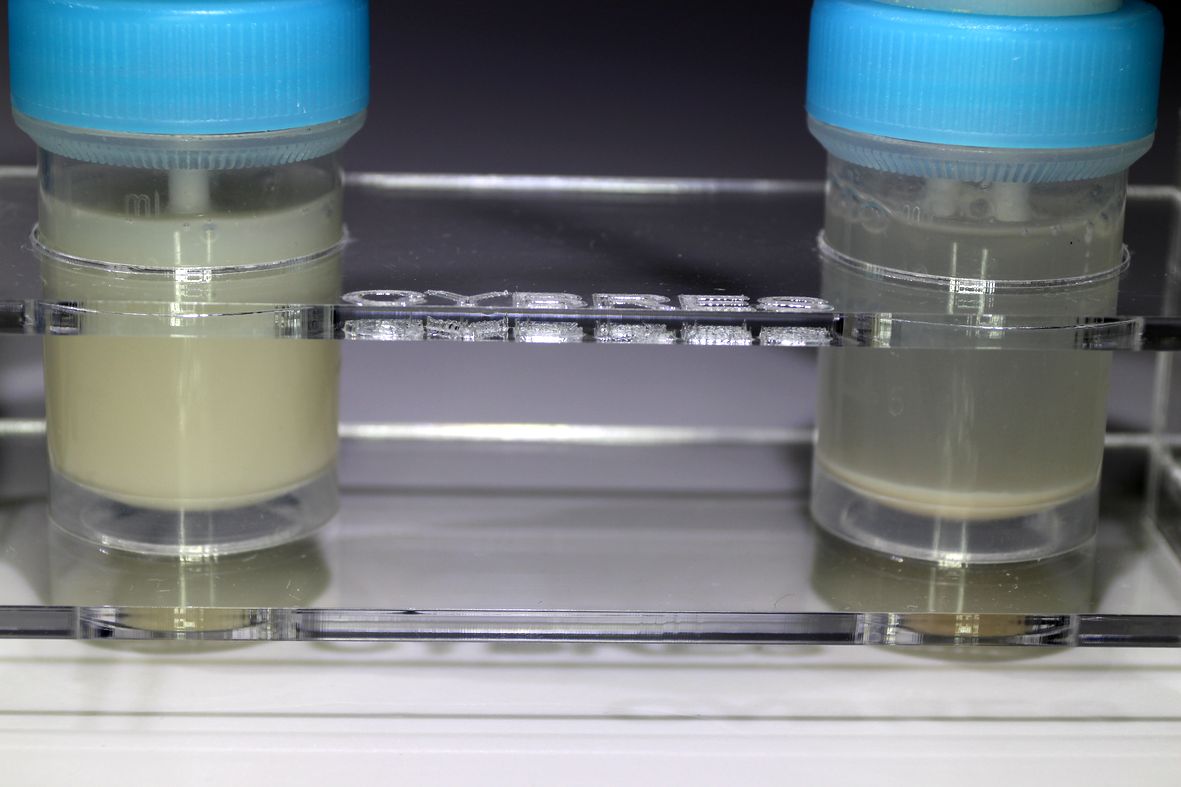}}
\caption{\small \textbf{(a)} Relation (as ch1/ch2 in \%) btw. impacted and non-impacted channels (origin is set to 0), data from Fig.\ref{fig:EIS}, the experiment 8.12, shown are 100Hz, 2.5kHz and 25kHz frequencies. Disturbances are generated by the fermentation activity of yeast;
\textbf{(b)} Sedimentation of yeast solution from original (left) to final (right) stages.\label{fig:EIS2}}
\end{figure}
The original solution for yeast has a number of different components, among them the ions $Zn^{+2}$, $Co^{+2}$, $Mg^{+2}$ and $Mn^{+2}$ that are metabolized by yeasts. This low dispersed solution represents in fact a colloidal fluid. The sedimentation process, see Fig. \ref{fig:sediment}, reduces ions and particles in the area between electrodes, it is reflected in decreasing the conductivity in the phase A. State of the art publications \cite{Kernbach13arXiv} point to the capability of WES to impact not only the activity of microorganisms, but also the rate of sedimentation, thus all three phases A, B, C can be used in sensing biological and physical impact of WES. Generally, the tested solutions point to the dependency
\begin{equation}
\label{eq:4}
\Delta C_B > \Delta C_C > \Delta C_A.
\end{equation}
As visible in Fig.~\ref{fig:EIS2}(a), the relative EIS dynamics has multiple disturbances that point to the ongoing yeast activities. In fact, the amplitude and appearance of disturbances can represent a EIS measurement parameter that characterizes the $CO_2$ production activity and can be measured as e.g. accumulated variance.

\begin{figure}[b!]
\centering
\subfigure{\includegraphics[width=.49\textwidth]{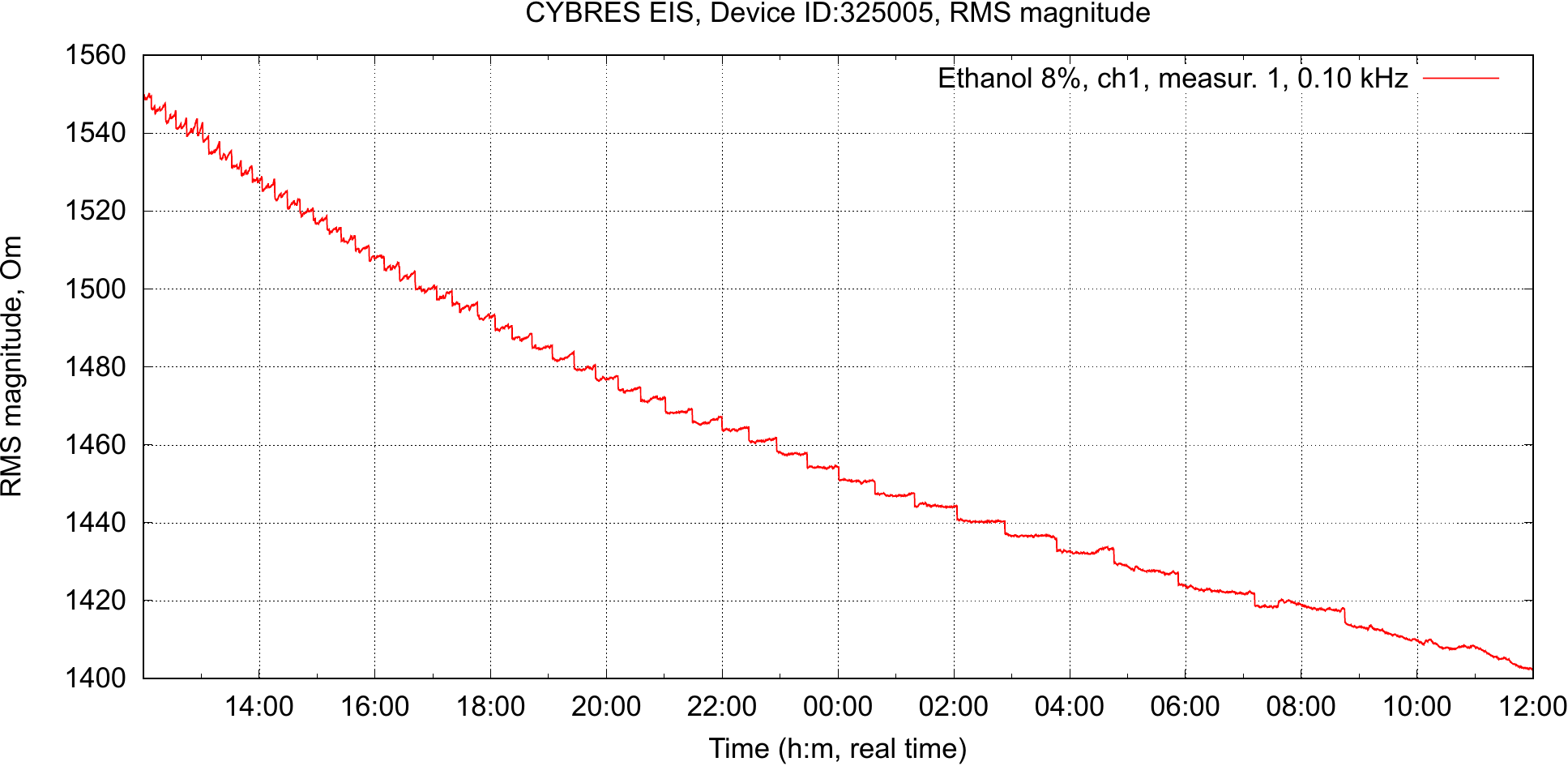}}
\subfigure{\includegraphics[width=.49\textwidth]{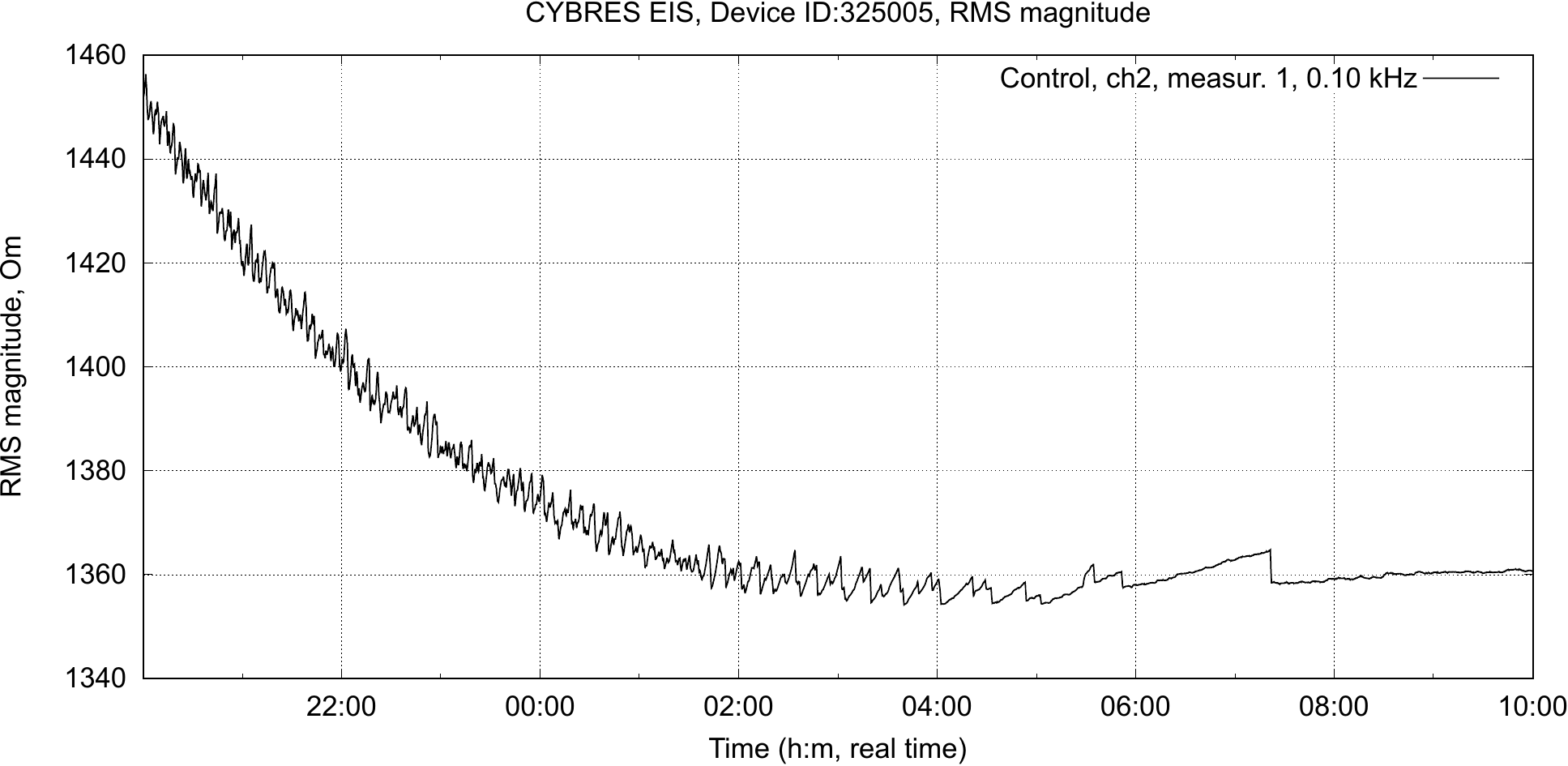}}
\caption{\small Oscillations of EIS dynamics at \SIlist{0.1}{\kilo\hertz}; \textbf{(a)} End of the fermentation process in the experimental channel with 8\% ethanol; \textbf{(b)} End of the fermentation process in the control channel. \label{fig:oscill}}
\end{figure}

Oscillations of metabolic dynamics in yeast are a well-known phenomenon, that occurs in actively growing cultures \cite{Chance1964}. The mechanism of oscillations is described in \cite{Madsen2005}, \cite{Bier2000} and shows the synchronization of glycolytic pathway between cells in the medium. The synchronization of metabolism is an evidence of the cell-cell communication and it is a key to the multicellular organisms development and survival \cite{Alberts2002}. The oscillations can be detected by different methods: $CO_2$ dynamics, pH level in the medium, fluorescence of the glycolytic intermediate NADH \cite{Laxman2010}. Male at al. demonstrated the possibility of using electrochemical methods for observation of metabolic cycles \cite{Male1999}. As shown in Fig. \ref{fig:oscill}, the impacted channels have different oscillation dynamics in comparison to the control channel, therefore oscillations in yeast can be concerned as one of the biosensor's informative parameters. The period of observed oscillations is \SIrange[range-phrase = --]{\approx 6}{ 11}{\minute}, what is consistent with the results of many studies \cite{Male1999}, \cite{Richard2003}, \cite{Bier2000}. Thus, concluding this section, a reliable approach for biological EIS measurements represents a combination of four following parameters:
\begin{itemize}
  \item the difference in amplitude of EIS dynamics in control and experimental containers. Important are not only absolute values but also long-term dynamics -- larger changes in EIS dynamics point to more active fermentation (i.e. ions exchange processes), see Fig. \ref{fig:ethdyn}(a);
  \item timing of different fermentation stages. Earlier begin points to faster (stimulated) activity. This parameter needs to be estimated manually based on EIS and temperature plots, e.g. start of disturbances, see Fig. \ref{fig:ethanol_long}.
  \item the level of disturbances (and to some extent oscillations) that characterizes the $CO_2$ production activity. It can be measured as a variance of the EIS dynamics calculated in a moving window and accumulated over the whole measurement, see Fig. \ref{fig:ethdyn}(b,d).
	\item the temperature of samples. Since the fermentation is an exothermic process and the thermostat set the equal external temperature on both samples, the channel with more intensive fermentation will produce more heat and thus its temperature is higher, see Fig. \ref{fig:ethanol_long}(b).
\end{itemize}

\section{Performed experiments}
\label{sec:experiments}

\subsection{Preparation of samples for pressure and EIS measurements, data representation}
\label{sec:preparation}

\begin{figure}[t!]
\centering
\subfigure[]{\includegraphics[width=.49\textwidth]{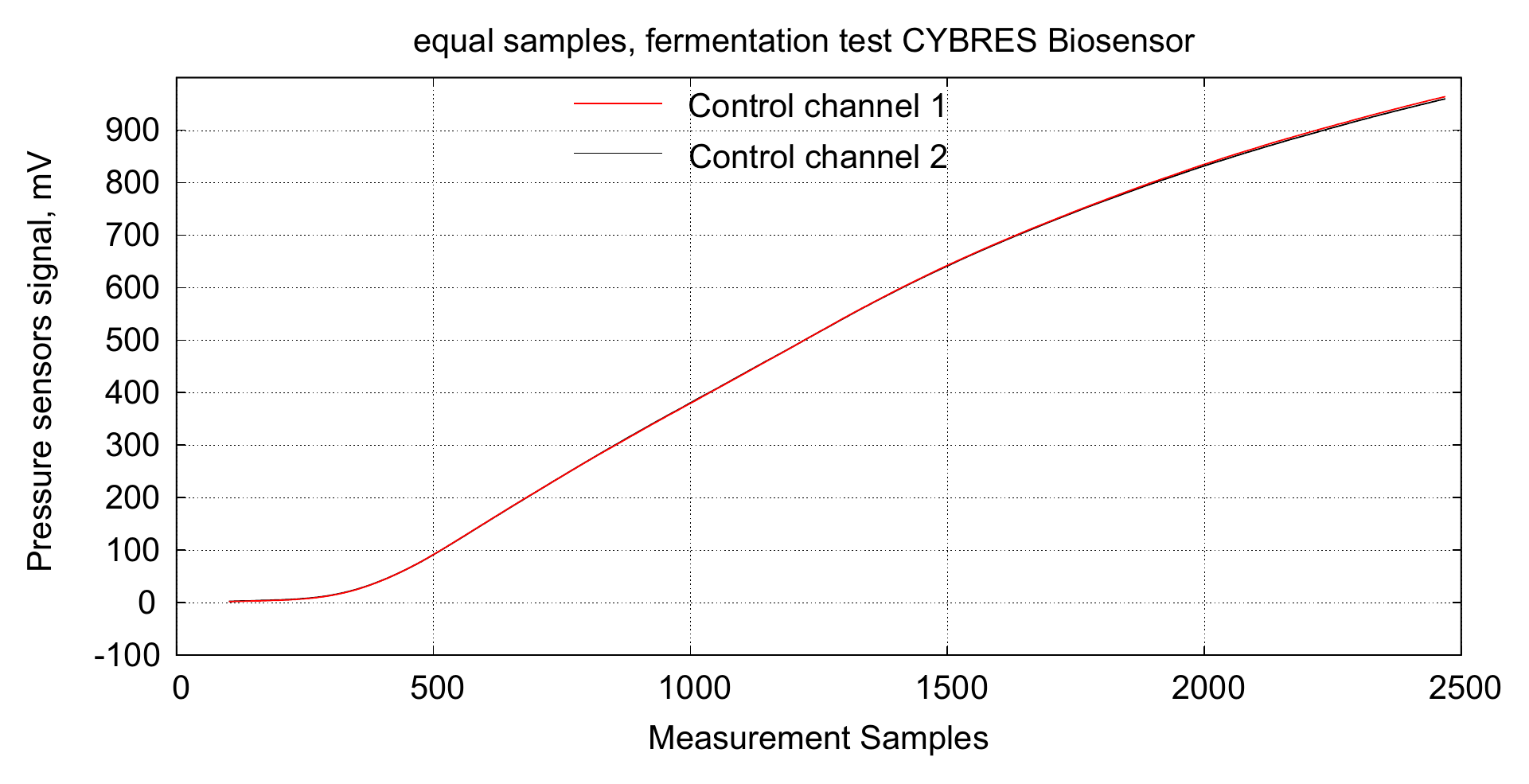}}
\subfigure[]{\includegraphics[width=.49\textwidth]{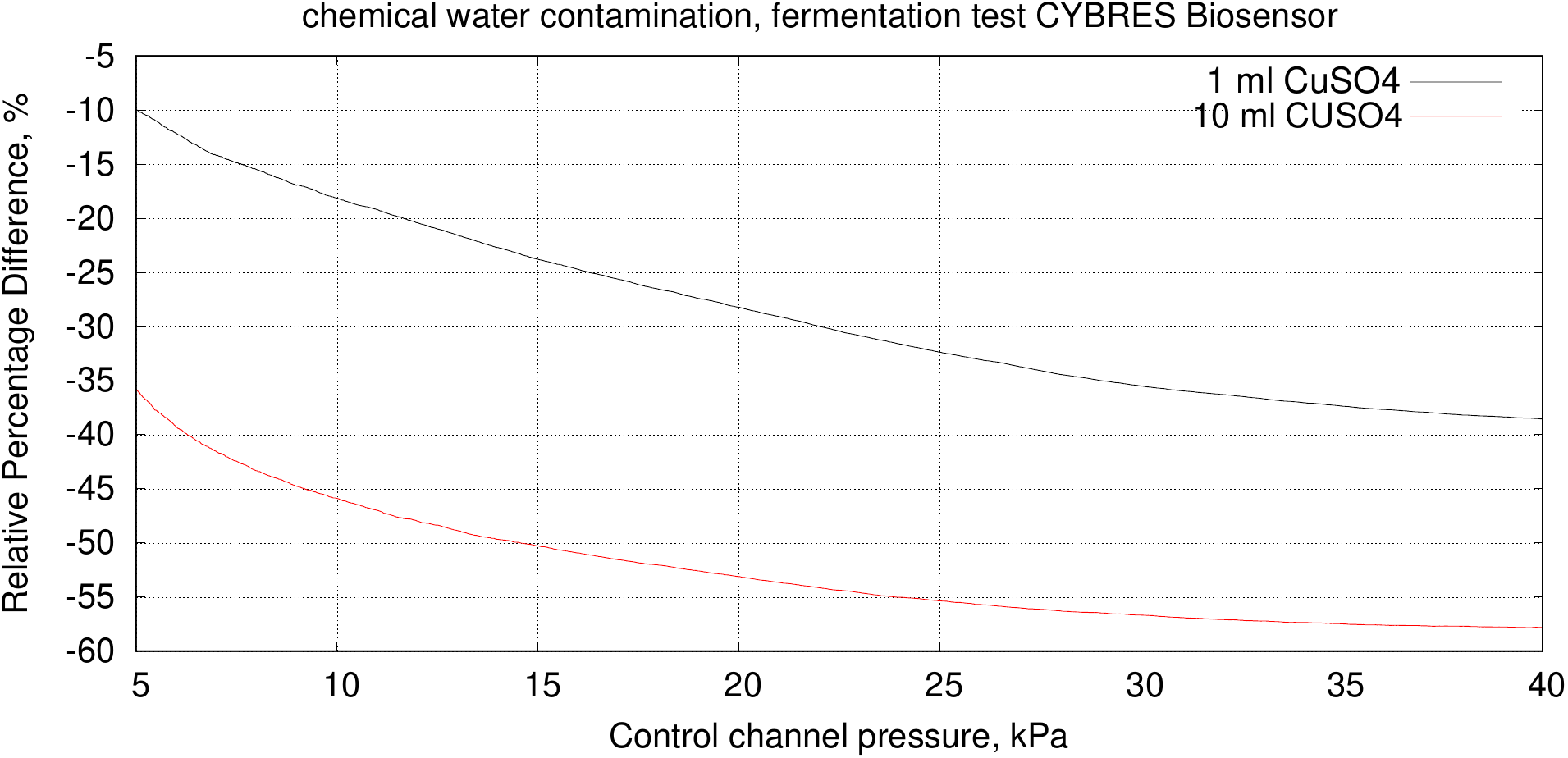}}
\subfigure[\label{fig:controlExperiments2}]{\includegraphics[width=0.45\textwidth]{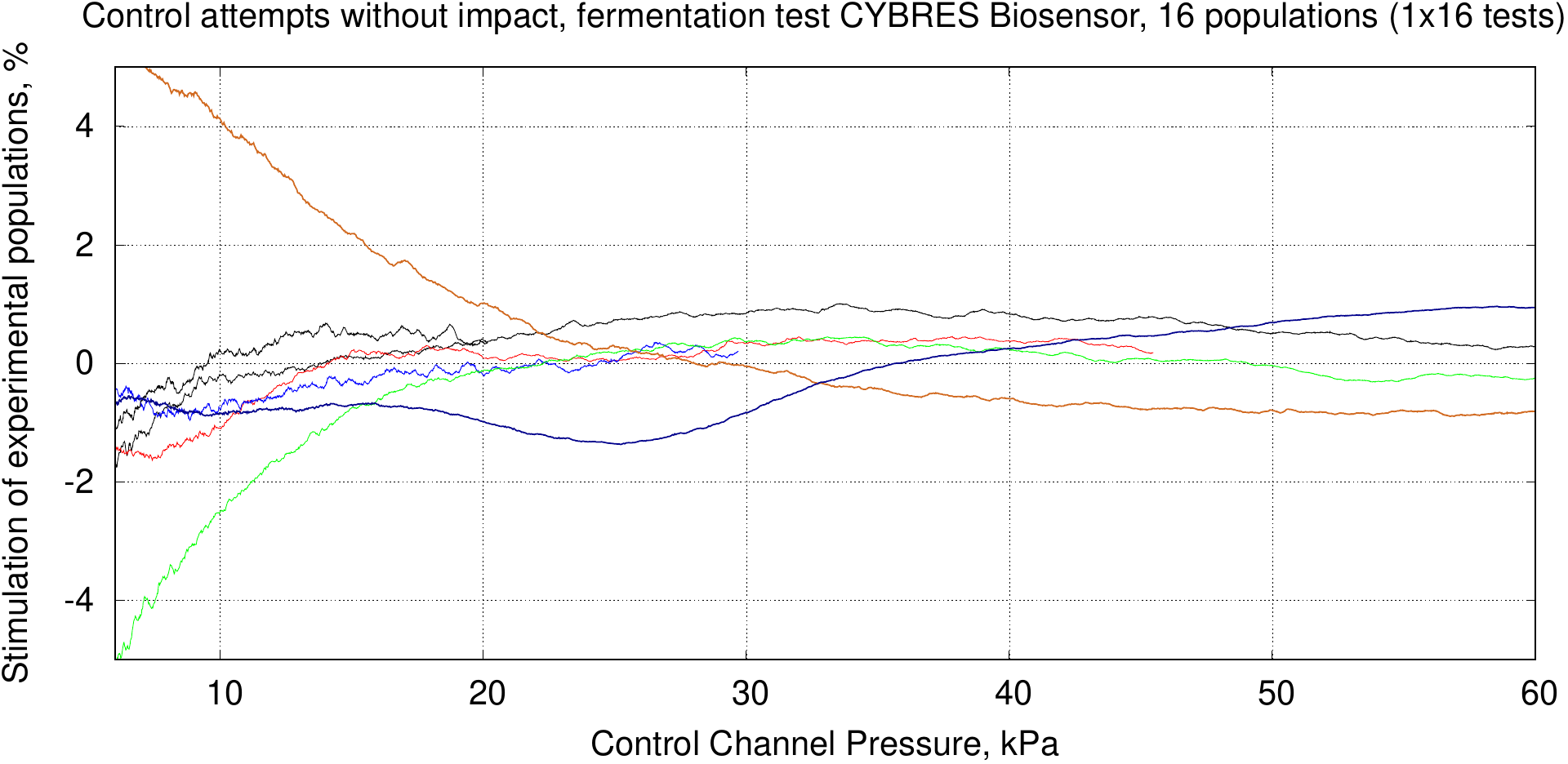}}
\caption{\small \textbf{(a)} Control attempt with equal conditions for both populations (equal samples); \textbf{(b)} Example of chemical water contamination -- $CuSO_4$ is added to the yeast solution in the experimental channel. Each curve represents an averaging of 8 pressure sensing channels, the x-axis represents measurement samples (about \SI{2}{\second}); \textbf{(c)} Repeated control measurements with equal conditions for both populations. \label{fig:controlExperiments}}
\end{figure}

All yeast samples, during exposition and preparation, were handled so that the temperature difference between samples was not larger than \SI{0.5}{\degreeCelsius} for \SI{3}{\minute}. Also the variations of sugar between probes, amount of water, amount and quality of yeast, variation of air pressure by closing the containers were kept as minimal as possible. It was achieved by dissolving all components in one large container and applying a high-resolution \SI{6}{\milli\litre} syringe for equal distribution of solution (\SI{5}{\milli\litre} were dispensed into each test tube) -- this provided the accuracy of \SI{0.1}{\milli\litre} or \SI{+-1}{\%}. The distilled water, dry bakery yeast (from the same manufacturer and from the same series) and lump sugar were used for experiments. Firstly, one cube of sugar (\SI{\approx3}{\gram}) was dissolved in \SI{200}{\milli\litre} of distilled water at room temperature and filled into test tubes. Tubes with sugar solution have been inserted into the devices, they were left undisturbed until the temperature stabilized with a tolerance of \SI{+-0.05}{\degreeCelsius}. The yeast solution was prepared in a similar way. A bag of dry bakery yeast (\SI{5}{\gram}) was dissolved in \SI{150}{\milli\litre} of distilled water at a room temperature until no yeast lumps were left. After this, 16 syringes for pressure sensing (or two syringes for EIS) were filled with this solution and inserted in pairs to the device - always one syringe to the control and experimental channel at a time. The yeast solution was stirred while being dispensed in the test tubes, this increased the homogeneity of solution and the repeatability of measurements.

Example of pressure measurement data for populations prepared in equal conditions is shown in Fig. \ref{fig:controlExperiments}(a), where each curve represents an averaging of 8 channels, and x-, y- axes are given by the number of samples and the output voltage of the sensor. This representation has several disadvantages, it demonstrates the temporal dynamics of fermentation in 'lag-' and 'log-' phases, but the results from different attempts are difficult to compare due to variation of this dynamics. More useful is another way to plot data, where the x-axis represents the pressure of control channel and the y-axis -- the relation of \emph{experimental/control} in \%, as shown in Fig. \ref{fig:controlExperiments2}. This method is used in all further pressure plots.

The yeast solution is inhomogeneous -- small clots of yeast that produce higher pressure in one or two channels and that increases a random inaccuracy of measurements. There are also technological reasons for pressure variations such as unequal heat distribution in the thermostat, and small deviations in preparation of 16 samples. For small variations of pressure and due to nonlinear fermentation dynamics, the software averaging of 'normal' and 'clotted' channels decreases an overall repeatability of measurements. These variations can be decreased by post-experimental data processing, e.g. by switching off $m$ 'abnormal' from all $N$ channels. This approach is used for small-signal pressure data processing (mainly for WES), for this the value of 'homogeneity of measurements' as m/N (in \%) is calculated. Typically, the homogeneity is on the level of \SIrange[range-phrase = --]{94}{96}{\%}, experiments with the homogeneity \SI{< 90}{\%} are discarded as invalid. The averaging for EIS measurements was performed in a 'physical way' by using large measurement containers (\SI{100}{\milli\litre} and more) to increase a homogeneity of yeast solution, see Fig. \ref{fig:LargeContainers}. To prevent active foam production we used anti-foaming agent 'Simeticon' \SI{0.5}{\milli\litre} per 200 ml of sugar solution. Since the temperature is the most influencing factor on yeast dynamics\cite{JONES1970} and pure water conductance\cite{Light2005}, it is essential to ensure equal conditions in both channels. A water bath with thermostabilization was used for that purpose. All water samples were placed into the water bath before yeast addition to exclude the unequal thermal influence.

\subsection{Control experiments}

\textbf{Control measurements for pressure sensors} included two types of tests: 1) all samples had equal solutions and the same environmental conditions during preparations and measurements; 2) \SIrange[range-phrase = --]{\approx 0.1}{1}{\%} of $CuSO_4$ was added to experimental samples. In this way the accuracy of measurement for equal samples and for strong chemical contamination can be estimated. Two examples of control measurements 1) and 2) are shown in Fig. \ref{fig:controlExperiments}, where each curve represents an averaging of 8 channels. As expected, we observe equal behavior of the test attempt 1) and inhibition of experimental channels (lower pressure) in the test attempt 2).
\begin{figure}[h!]
\centering
\subfigure{\includegraphics[width=.49\textwidth]{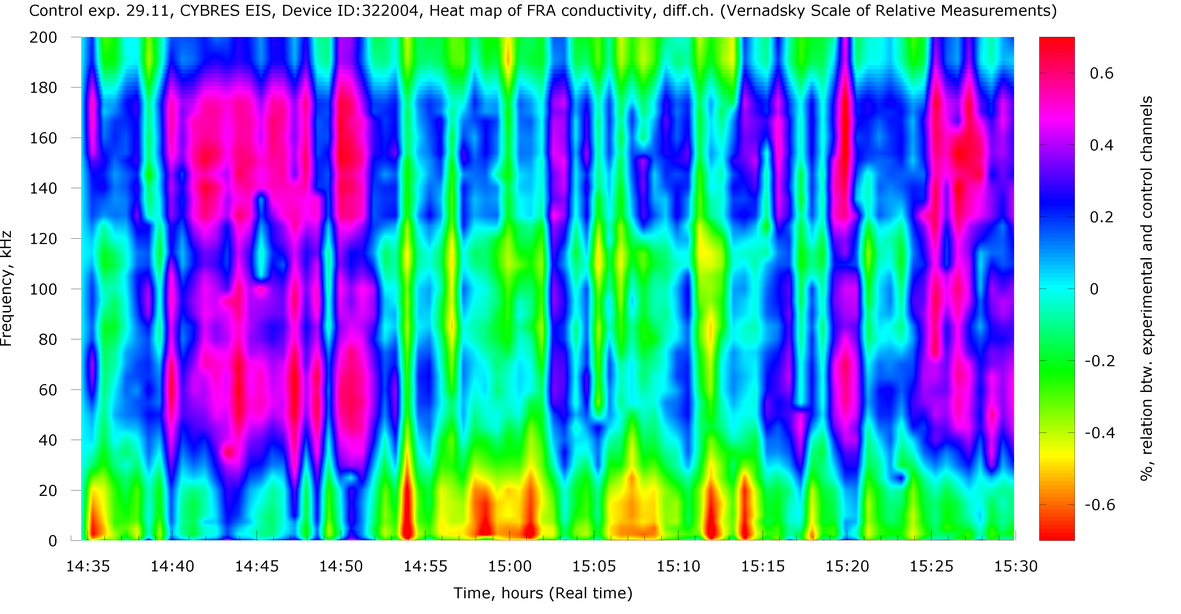}}
\subfigure{\includegraphics[width=.49\textwidth]{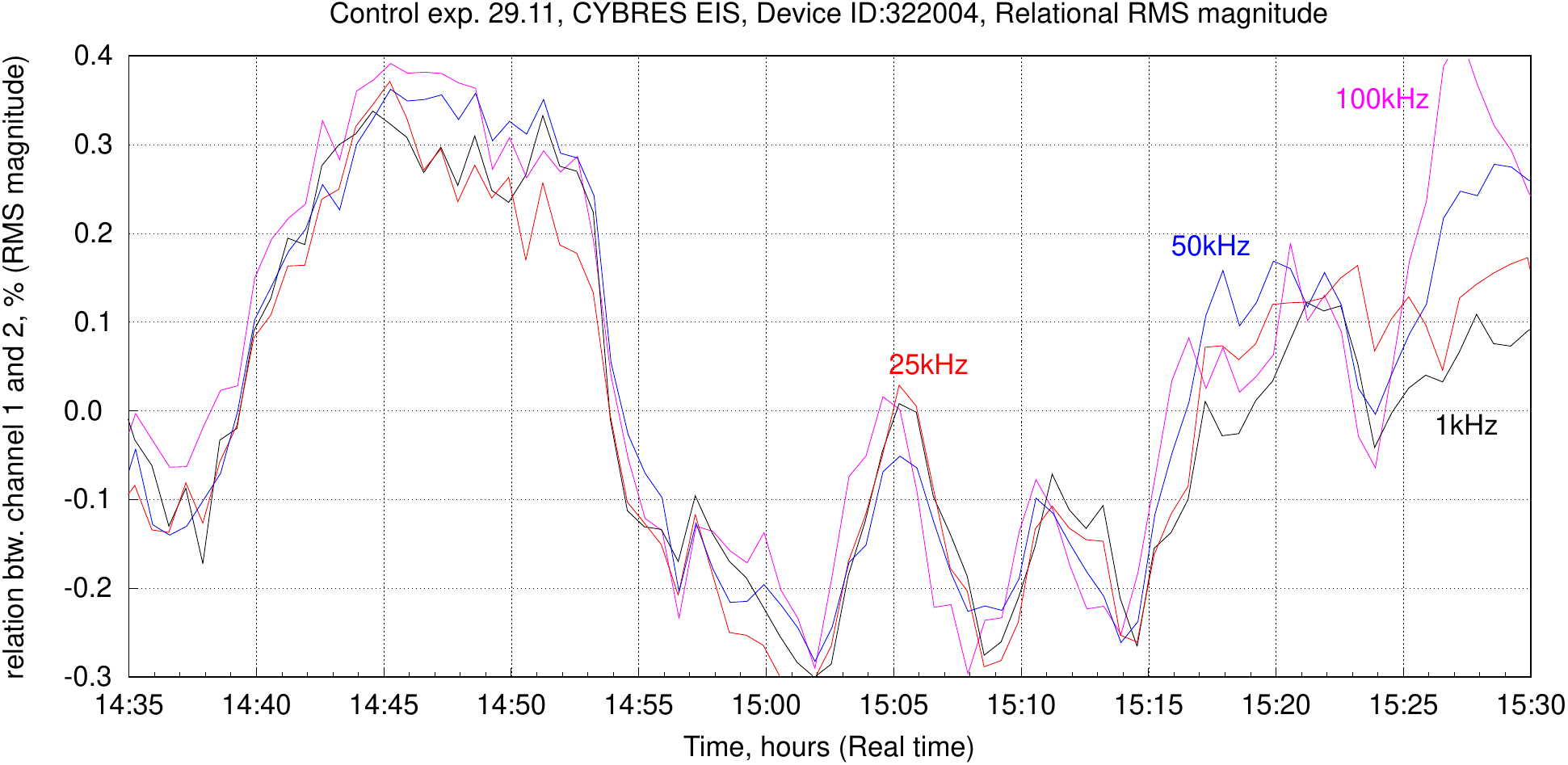}}
\caption{\small \textbf{(a)} EIS dynamics of the control attempt without any impact, the reference point is selected in the way shown in Fig. \ref{fig:EIS}(a); \textbf{(b)} 2D plot at \SIrange[range-phrase = --]{1}{100}{\kilo\hertz}, the variation is about \SI{+-0.4}{\%} (origin shifted to zero). \label{fig:controlEIS}}
\end{figure}
Control measurements with equal conditions for both populations were repeated 7 times with 16 pairs of samples, in total 112 samples, the homogeneity of measurements is \SI{93,75}{\%}, see Fig. \ref{fig:controlExperiments2}. We observe a higher variation of results at lower pressure that is explained firstly by inaccuracy of sensors for low-range measurements, secondly, by the variation of fermentation dynamics in the initial 'lag-'phase. Generally, for pressures \SI{> 30}{\kilo\pascal} we expect about \SI{+-1}{\%} of inaccuracy to repeated measurements that corresponds to the estimated inaccuracy of filing the yeast solution with about \SI{+-1}{\%}. At higher pressure the fermentation dynamics is inhibited due to higher concentration of ethanol. Thus, the values in middle range, at e.g. \SIrange[range-phrase = --]{25}{30}{\kilo\pascal}, can be used as a single-value-characterization of the dynamics.

\begin{figure*}[ht]
\centering
\subfigure[]{\includegraphics[width=.49\textwidth]{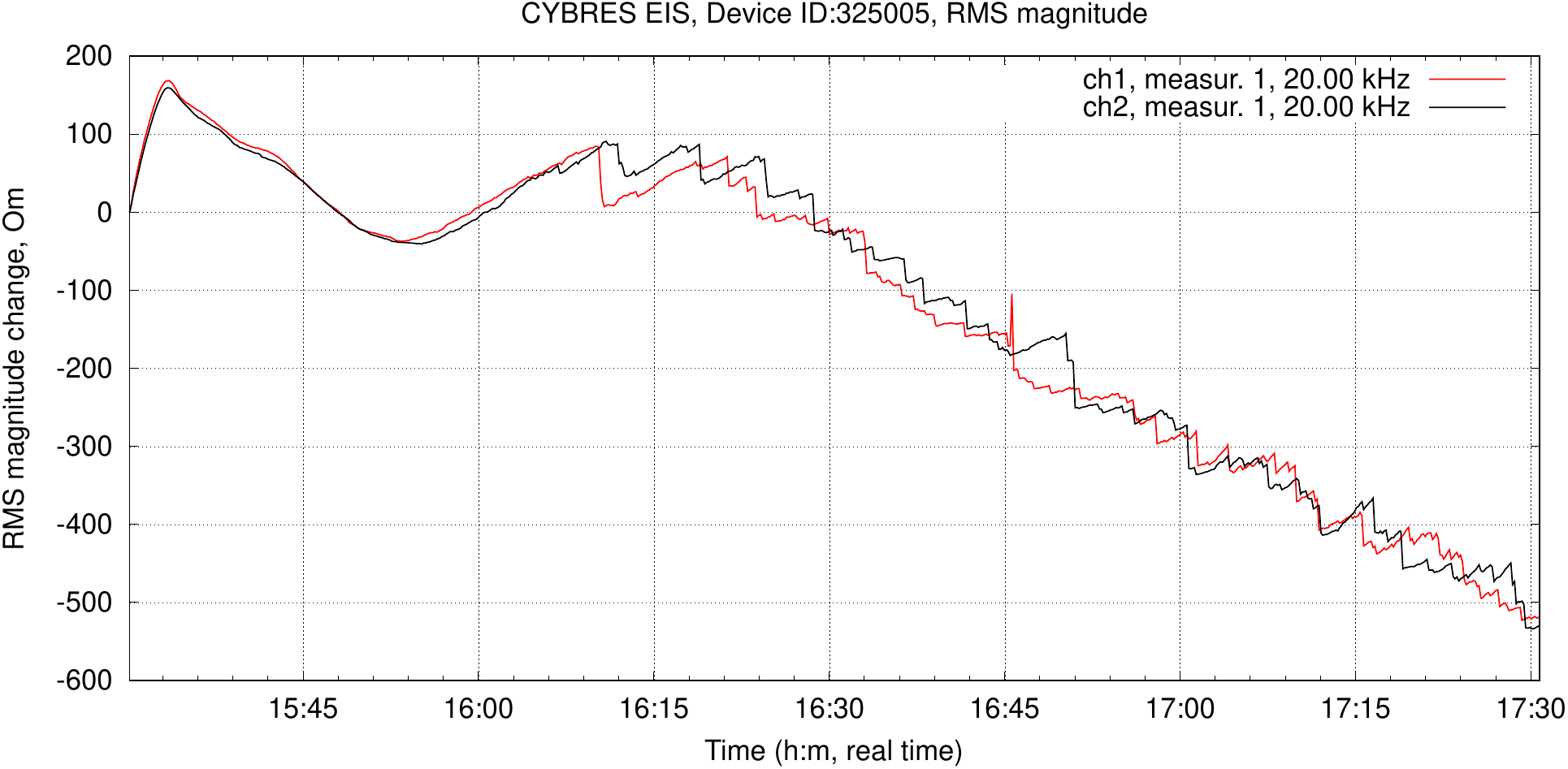}}~
\subfigure[]{\includegraphics[width=.49\textwidth]{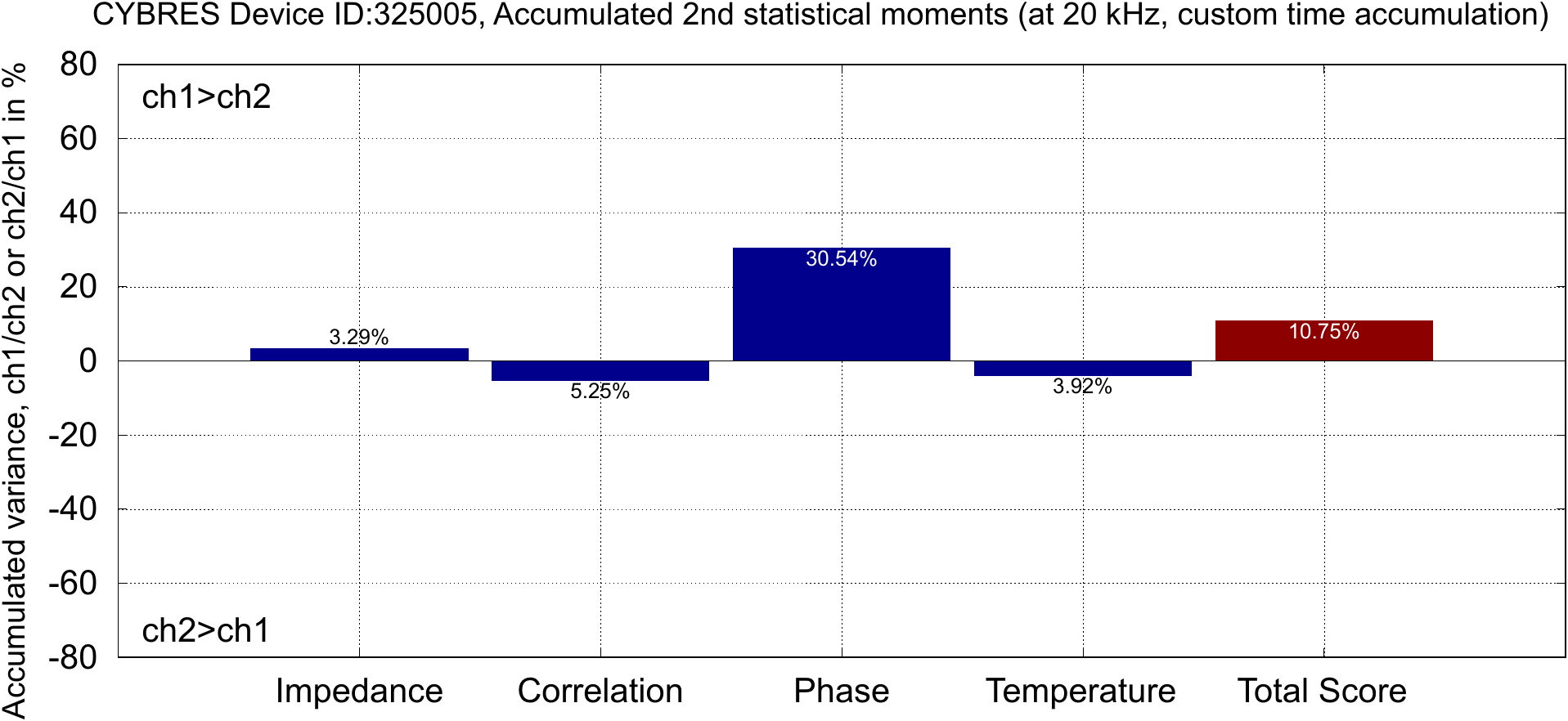}}
\subfigure[]{\includegraphics[width=.49\textwidth]{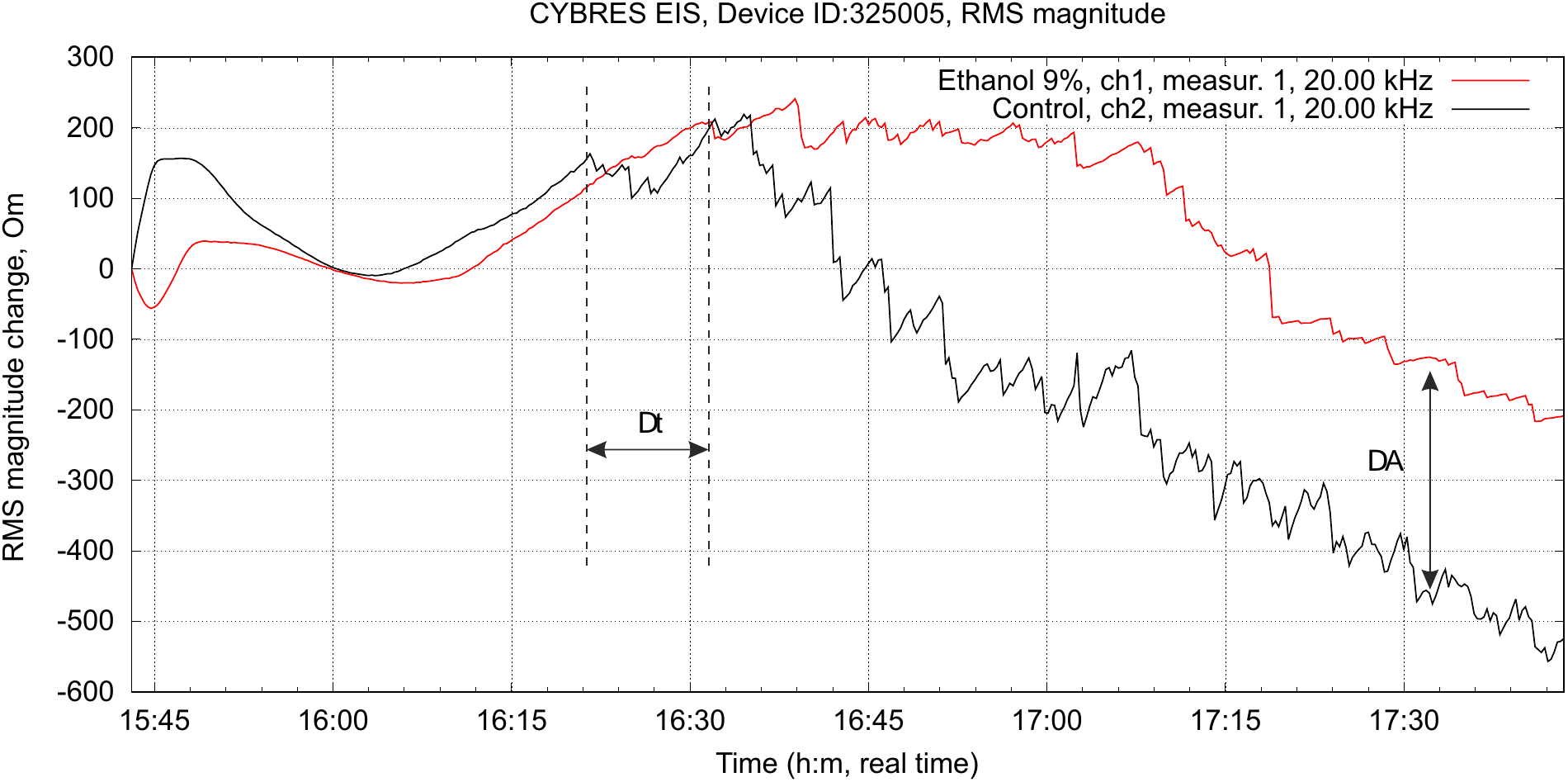}}~
\subfigure[]{\includegraphics[width=.49\textwidth]{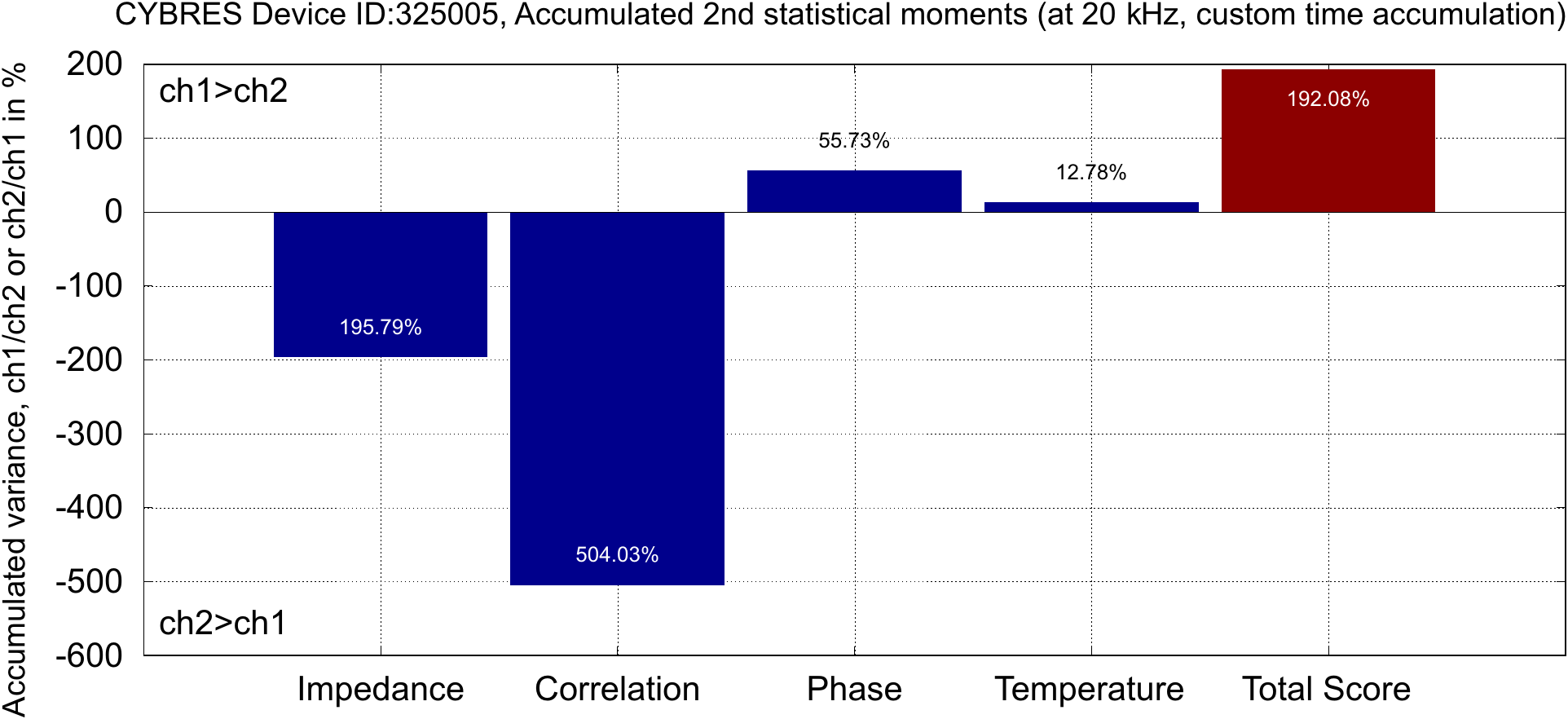}}
\caption{\small Yeast fermentation EIS dynamics at \SIlist{20}{\kilo\hertz}. Control measurement: \textbf{(a)} RMS magnitude; \textbf{(b)} Variation accumulated for the whole run. Experimental attempt with 9\% Ethanol in Ch1: \textbf{(c)} RMS magnitude; \textbf{(d)} Accumulated variance.  \label{fig:condyn} \label{fig:ethdyn}}
\end{figure*}

\textbf{Control measurements for EIS sensors} are shown in Figs. \ref{fig:controlEIS} and \ref{fig:ethdyn}(a,b). The behaviour in Fig. \ref{fig:controlEIS} demonstrates the noised dynamics with amplitude of \SI{+-0.4}{\%}--\SI{+-0.5}{\%} (for \SI{15}{\milli\litre} containers) after the reference point, selected in the way shown in Fig. \ref{fig:EIS}(a). The impact of different frequencies on the noise production is on the level of \SI{+-0.1}{\%}. Figure \ref{fig:ethdyn} shows experiments with 100 ml containers. Here the disturbances of EIS dynamics produced by $CO_2$ generation are calculated as a variance in a moving window (90 samples) and then it is accumulated for the whole run. In order to avoid initial variation of parameters (buffering the sliding window, inequality of yeast in each pack, variation of mixing time and environmental conditions), we recommend to analyze the first hour of the fermentation, started 5 minutes before the begin of fermentation process (based on temperature curves). Figure \ref{fig:ethdyn}(a,b) shows a control attempt, and Fig. \ref{fig:ethdyn}(c,d) the experiment with impaired yeast by ethanol addition (9\%) -- this experiment demonstrates an inhibition effect. We observe a clear difference in amplitude of EIS dynamics ($\Delta A$), in time when the disturbances started (the phase difference $\Delta t$) as well as the accumulated variance. The 'Total Score' (the sum of absolute values of all bars) -- the difference of accumulated variance between control and inhibited channels is about 193\%. The result of 21 control attempts is shown in Table \ref{tab:ControlTableFermentation}, the 'Total Score' parameter  is \num[separate-uncertainty]{18.71(783)}{\%}. Thus, we observe about 9x-10x difference between 'equal conditions for both populations' and 'one population is inhibited by 9\% of ethanol'.

\begin{figure}[h!]
\centering
\subfigure[]{\includegraphics[width=.49\textwidth]{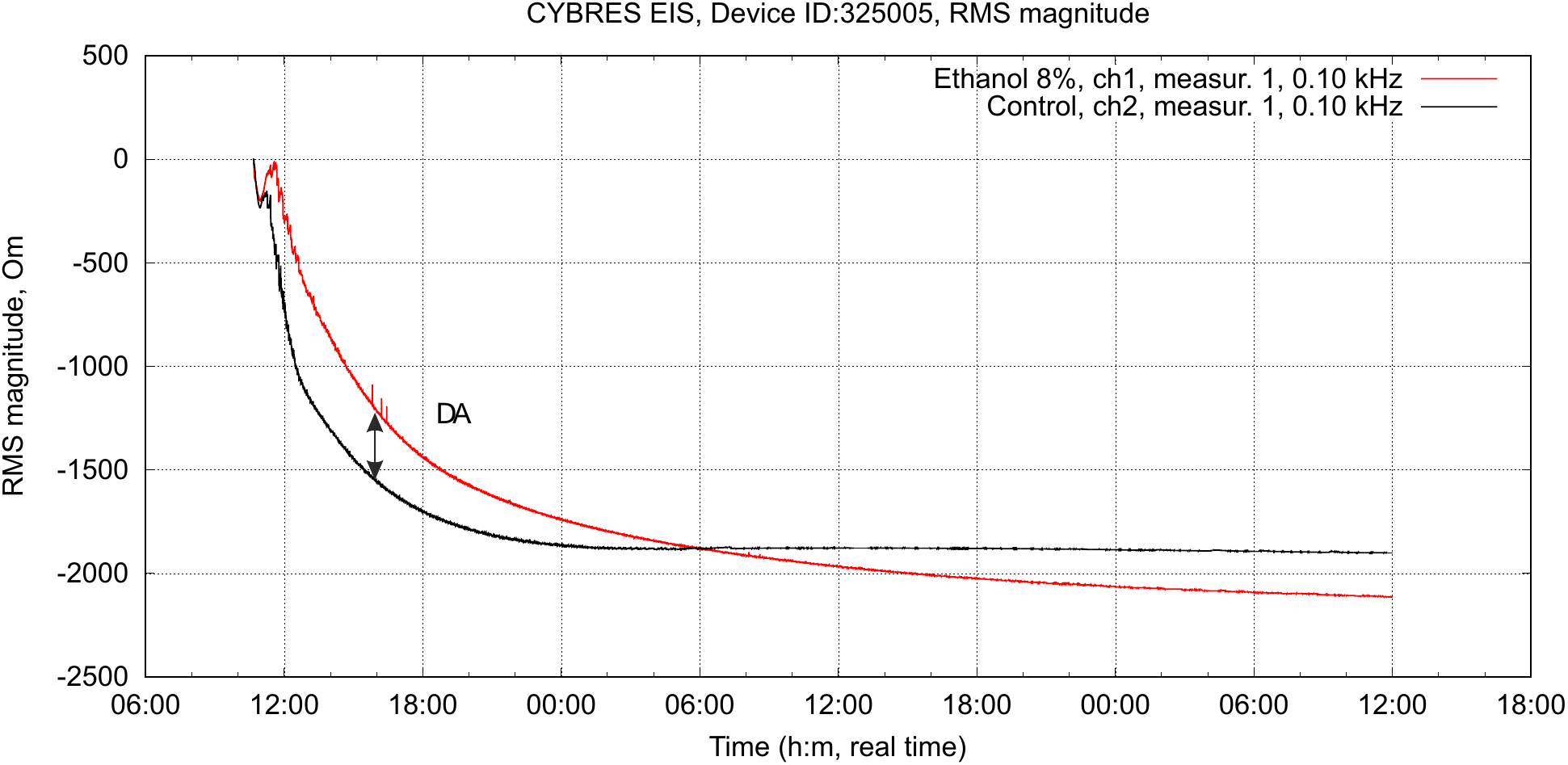}}
\subfigure[]{\includegraphics[width=.49\textwidth]{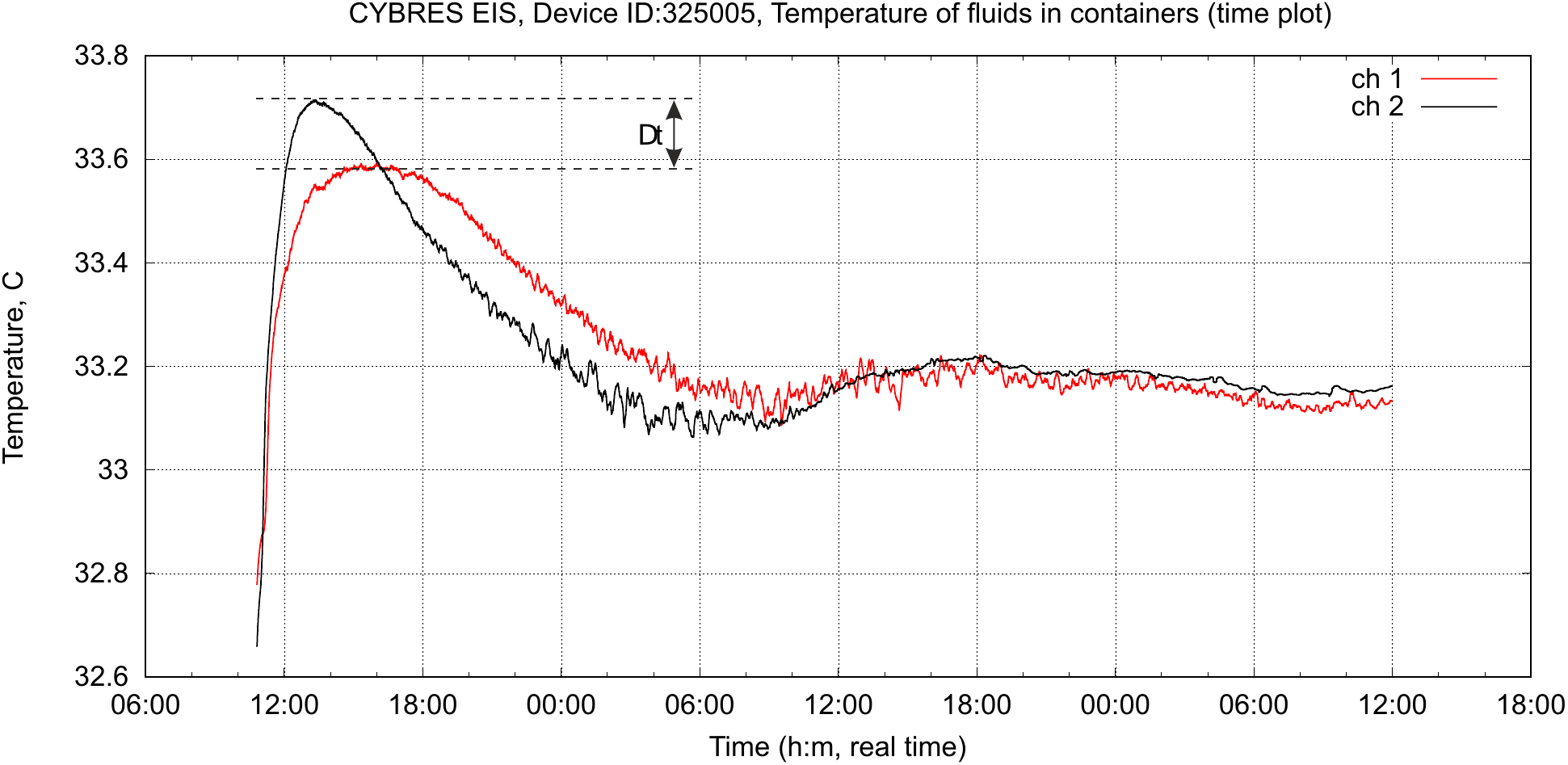}}
\caption{\small Long-term yeast fermentation EIS dynamics at \SIlist{0.1}{\kilo\hertz} with 8\% Ethanol in Ch1. \textbf{(a)} RMS magnitude; \textbf{(b)} Fluids temperature in containers. \label{fig:ethanol_long}}
\end{figure}

Long-term yeast fermentation EIS dynamics with 8\% Ethanol in Ch1 is shown in Fig. \ref{fig:ethanol_long}. Similarly the the previous measurement, we observe here the difference in amplitude $\Delta A$ of EIS dynamics and the difference in temperature (the temperature difference $\Delta t$) -- the channel 1 with inhibited yeasts produce less heat and has a lower temperature during the initial stage of fermentation.

\begin{table*}[ht]
\centering
\caption{\small Control measurements of two unprocessed distilled water with EIS approach (100 ml containers). Measurements are performed within 60 min. after begin of the fermentation process as estimated by the temperature curves. \label{tab:ControlTableFermentation}}
\fontsize {9} {10} \selectfont
\begin{tabular}{
p{2.0cm}@{\extracolsep{3mm}}
p{2.0cm}@{\extracolsep{3mm}}
p{2.0cm}@{\extracolsep{3mm}}
p{2.0cm}@{\extracolsep{3mm}}
p{2.0cm}@{\extracolsep{3mm}}
p{2.0cm}@{\extracolsep{3mm}}
}\hline
   & Impedance,\%  & Correlation,\% & Phase,\% & Temperature,\% & Total Score,\%   \\\hline
$\mu \pm \sigma$  &  \num[separate-uncertainty]{23.32(1586)}  &  \num[separate-uncertainty]{26.24(2043)}  &  \num[separate-uncertainty]{14.39(990)}  &  \num[separate-uncertainty]{10.89(902)}  &  \num[separate-uncertainty]{18.71(783)}   \\
\hline
\end{tabular}
\end{table*}

\subsection{Assessing the quality of water}
\label{sec:distantWES}

\begin{figure}[h!]
\centering
\subfigure{\includegraphics[width=.49\textwidth]{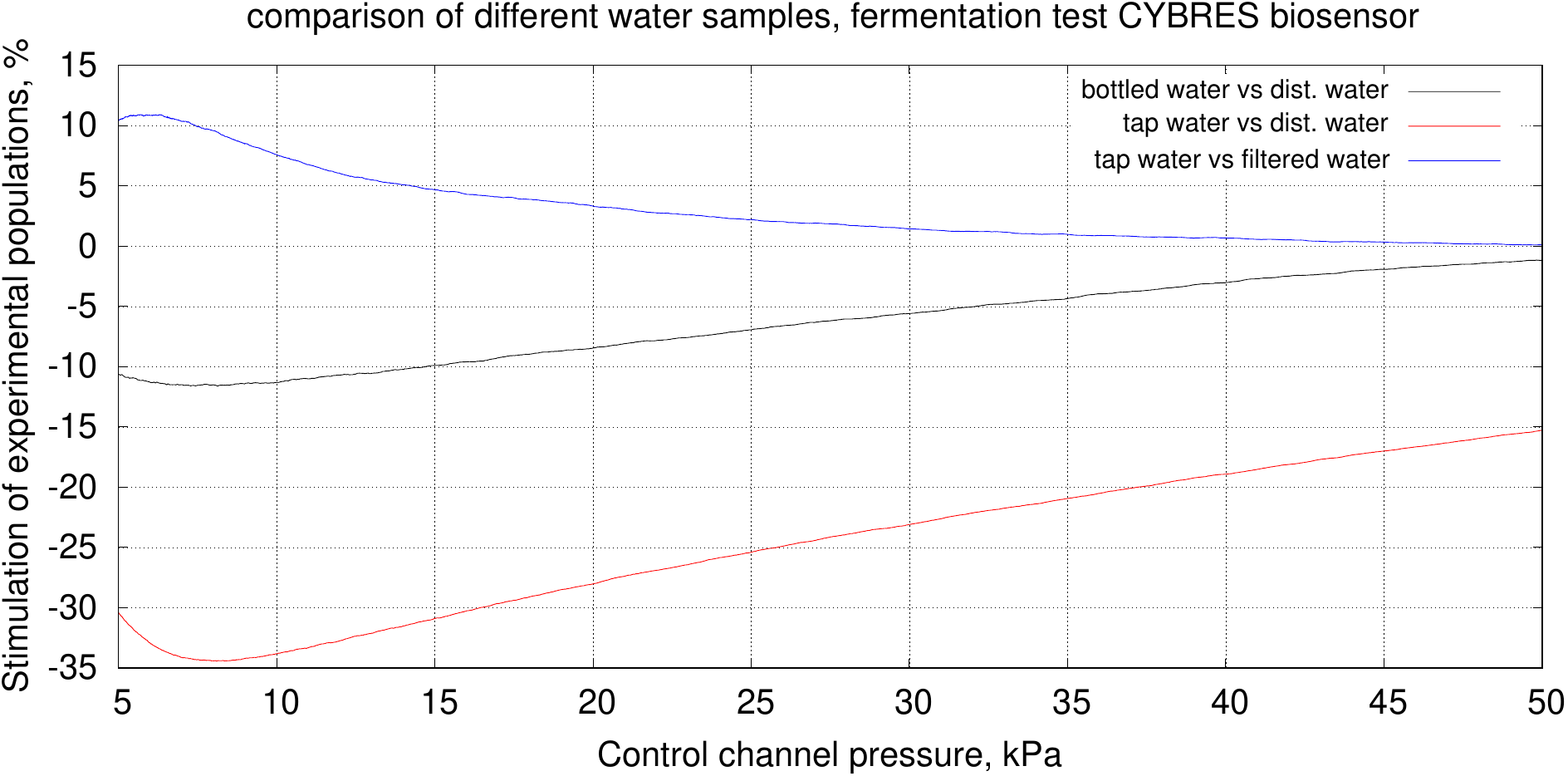}}
\caption{\small Biological measurements of water quality: comparing the tap, bottled and filtered water with the reference to distilled water (measured by the pressure sensing approach). \label{fig:filters}}
\end{figure}

The quality of water, incl. drinking water, is measured by yeast in biological way, i.e. how the particular water sample stimulates or inhibits the metabolism of microorganisms. Due to differential approach, it is possible to measure one sample always in comparison to another one. Here we face the problem of selecting some reference samples, it makes sense to select the pure distilled water for such a reference. Fig. \ref{fig:filters} shows several examples of comparing the tap, bottled and filtered water with the reference to distilled water. Especially interesting are results related to the tap water that demonstrate a strong inhibition of microorganisms (even in relation to the bottled water).

\subsection{Interpopulation biophotonic \SI{>320}{\nano\meter} communication}

\begin{figure}[h!]
\centering
\subfigure{\includegraphics[width=.49\textwidth]{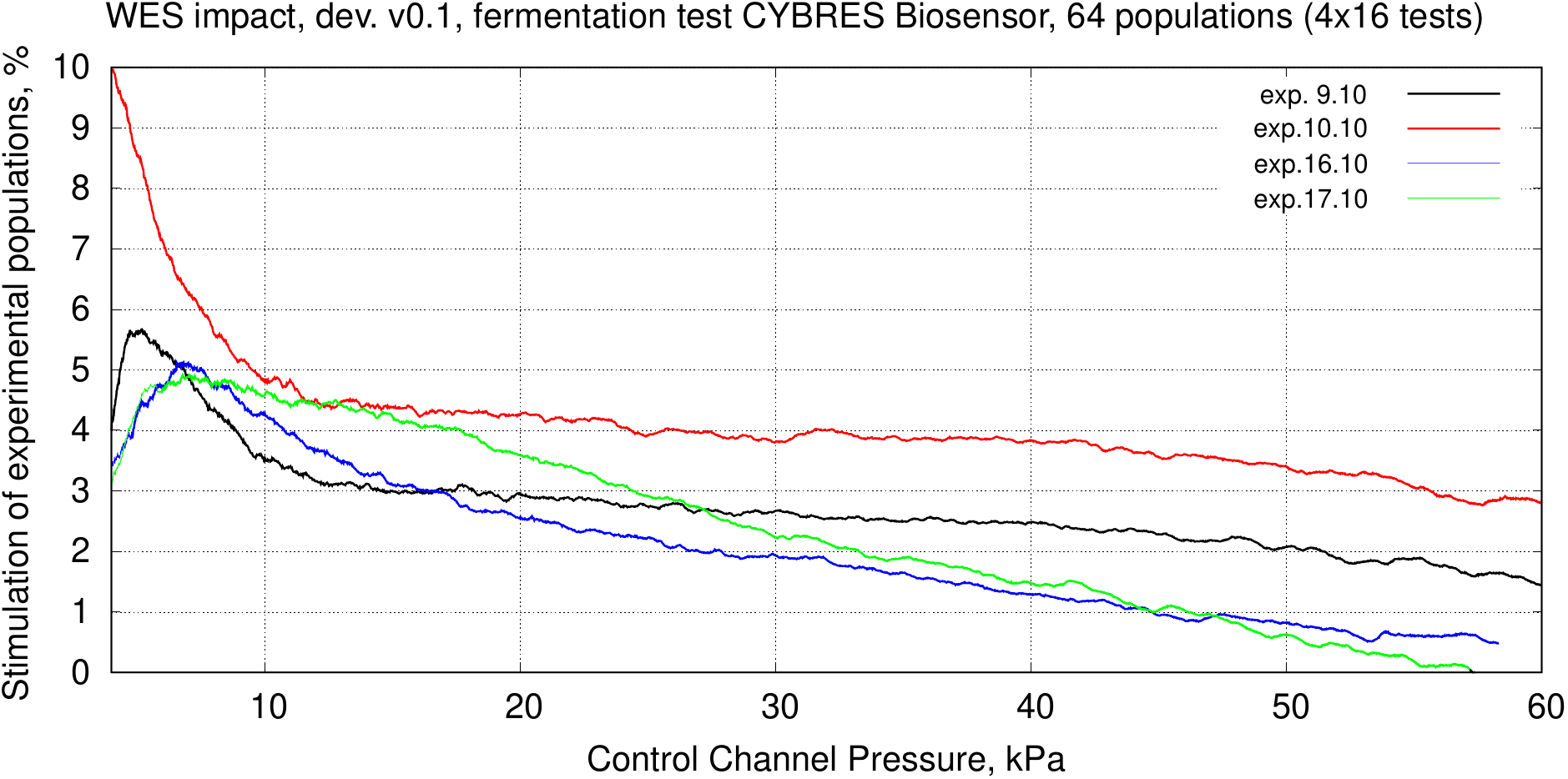}}
\caption{\small Four repeating experiments with stimulation of fermentation activity (wavelength \SI{> 320}{\nano\meter}) between two populations of yeasts, see description in text (measured by the pressure sensing approach). \label{fig:WES1}}
\end{figure}

We replicated the well-known experiments with the biophotonic emission \cite{Gurvich44en}, \cite{Popp94}, \cite{Kaznacheev81en} between two populations of yeasts. In particular, we closely followed the Kaznacheev's experiment -- the setup from \cite{Kaznacheev81en} with two populations in separate containers, where an initiated inhibition of one population leaded to a spontaneous inhibition of another population. We implemented the 'stimulated version' of the Kaznacheev's experiment -- the experimental samples without sugar were placed closely to the population of yeast with sugar (the distance -- about \SI{10}{\milli\meter}), where the fermentation had already started. Both populations were stored in glass/PP containers that absorbers UV part of spectra with wavelength less than \SI{\approx 320}{\nano\meter}. The exposition time was about \SIrange[range-phrase = --]{3}{4}{\hour}, since the volumes of both containers are comparable to each other, the temperature difference between control and experimental samples during exposition was neglectfully small. Four performed attempts are shown in Fig. \ref{fig:WES1}. We observe a weak stimulation about \SI{2.75}{\%} in average in all experimental attempts. 

\begin{figure}[h!]
\centering
\subfigure[\label{fig:LargeContainers}]{\includegraphics[width=.35\textwidth]{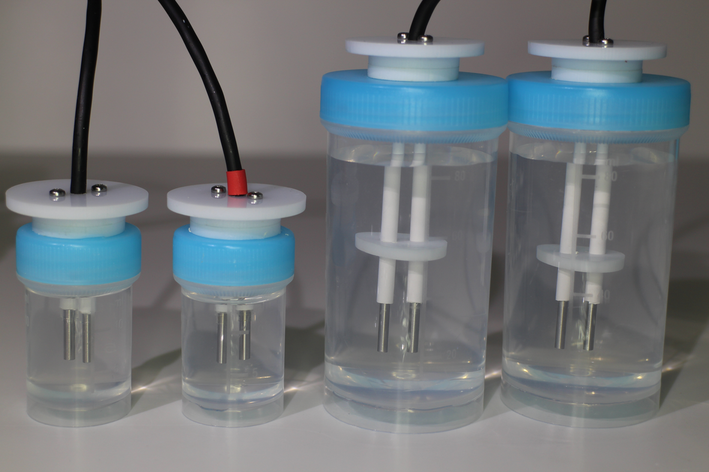}}
\subfigure[\label{fig:LargeContainersDynamics}]{\includegraphics[width=.5\textwidth]{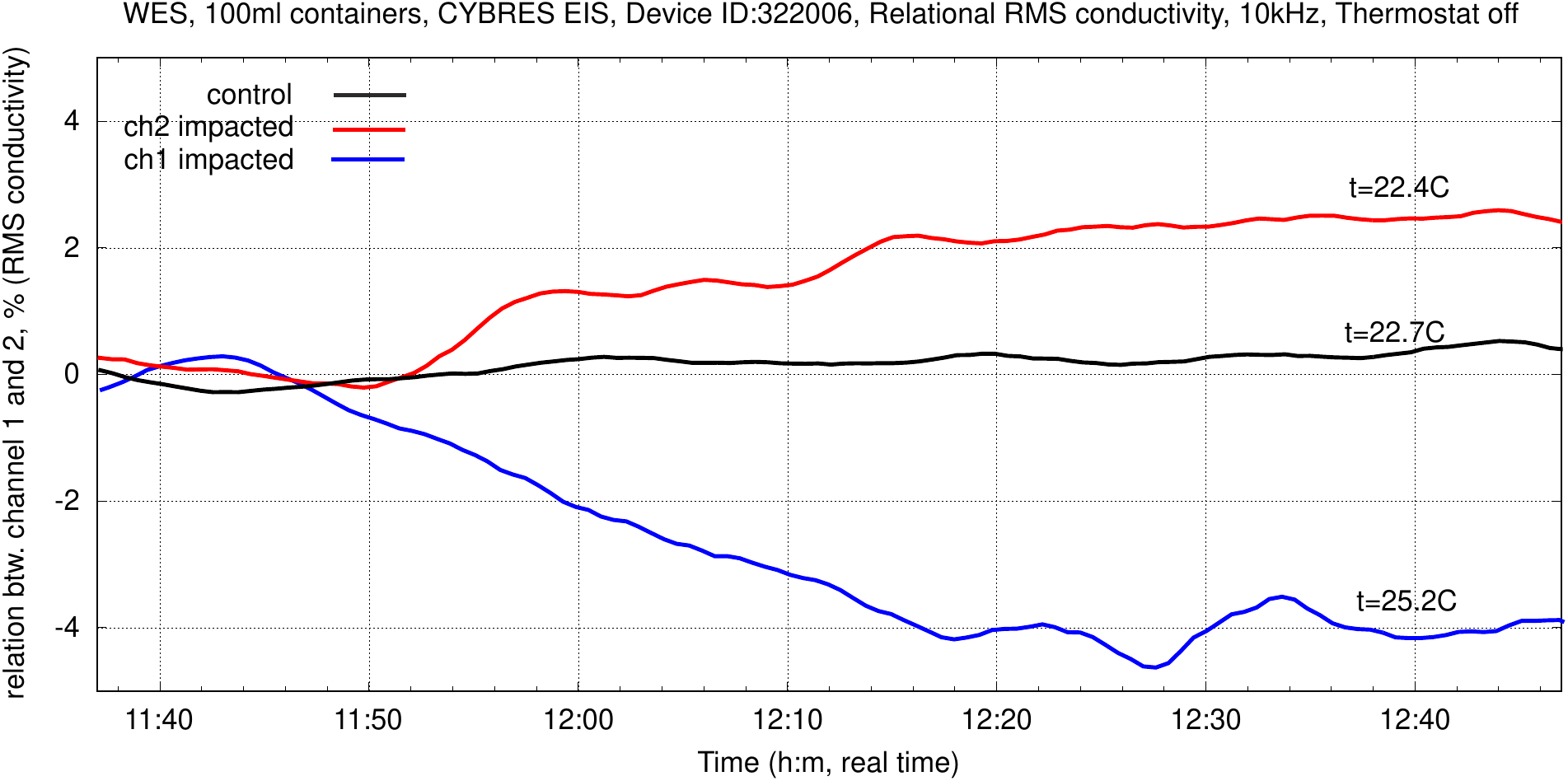}}
\caption{\small Replication of results shown in Fig. \ref{fig:WES1} by the EIS approach: \textbf{(a)} Large \SI{100}{\milli\litre} containers for EIS attempts; \textbf{(b)} EIS dynamics with \SI{100}{\milli\litre} containers of three attempts -- control, experimental ch.1 and ch.2 -- with WES (see description in text).}
\end{figure}

These attempts with some minor modifications were replicated with the EIS approach. To increase the homogeneity of yeast solution, large containers with \SI{100}{\milli\litre}, see Fig. \ref{fig:LargeContainers}, were utilized. In order to avoid possible technological artefacts, first the channel 1 and then the channel 2 were selected as experimental channels (i.e. we expected a symmetric inverse dynamics of both attempts). The comparison of these and one control measurements are shown in Fig. \ref{fig:LargeContainersDynamics}. Results indeed demonstrated the inverse dynamics of both experiments, the unsymmetrical shape can be explained by a small temperature variation during exposition. Thus, the EIS data confirm the pressure measurements and state-of-the-art-publications on weak effects of distant interactions between separate microbiological populations.

\begin{figure*}[htp]
\centering
\subfigure[]{\includegraphics[width=.49\textwidth]{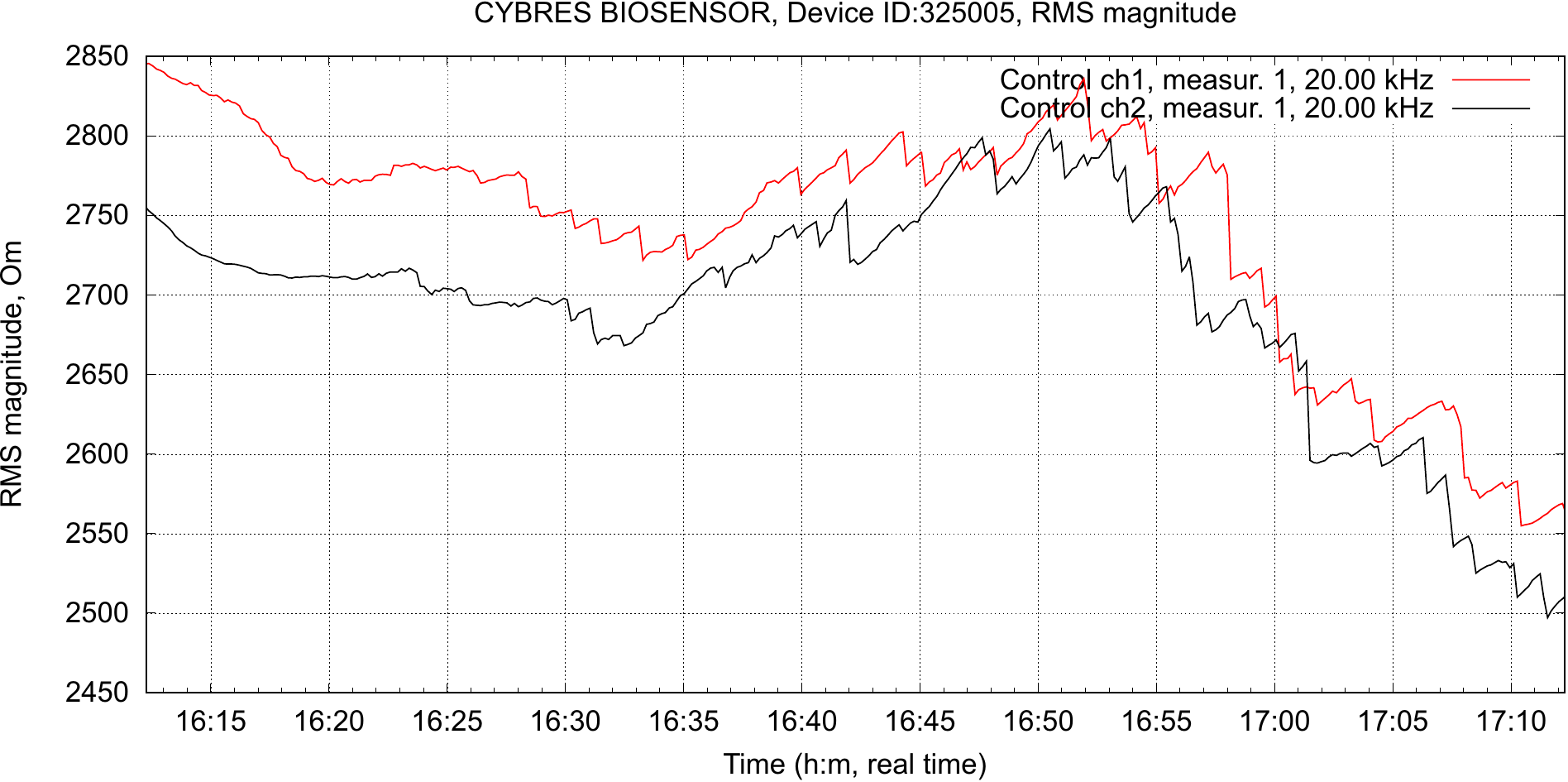}}~
\subfigure[]{\includegraphics[width=.49\textwidth]{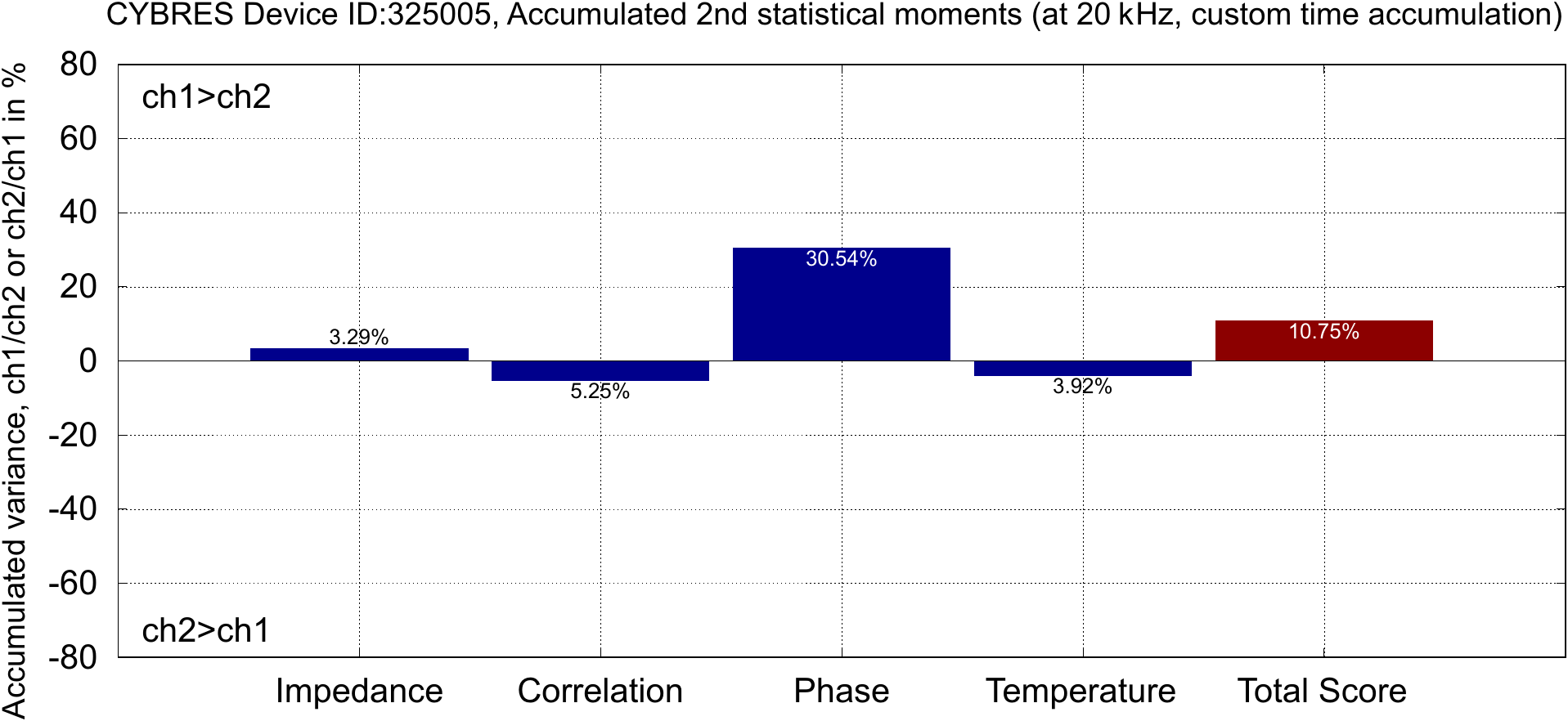}}
\subfigure[\label{fig:MEMONdyn1}]{\includegraphics[width=.49\textwidth]{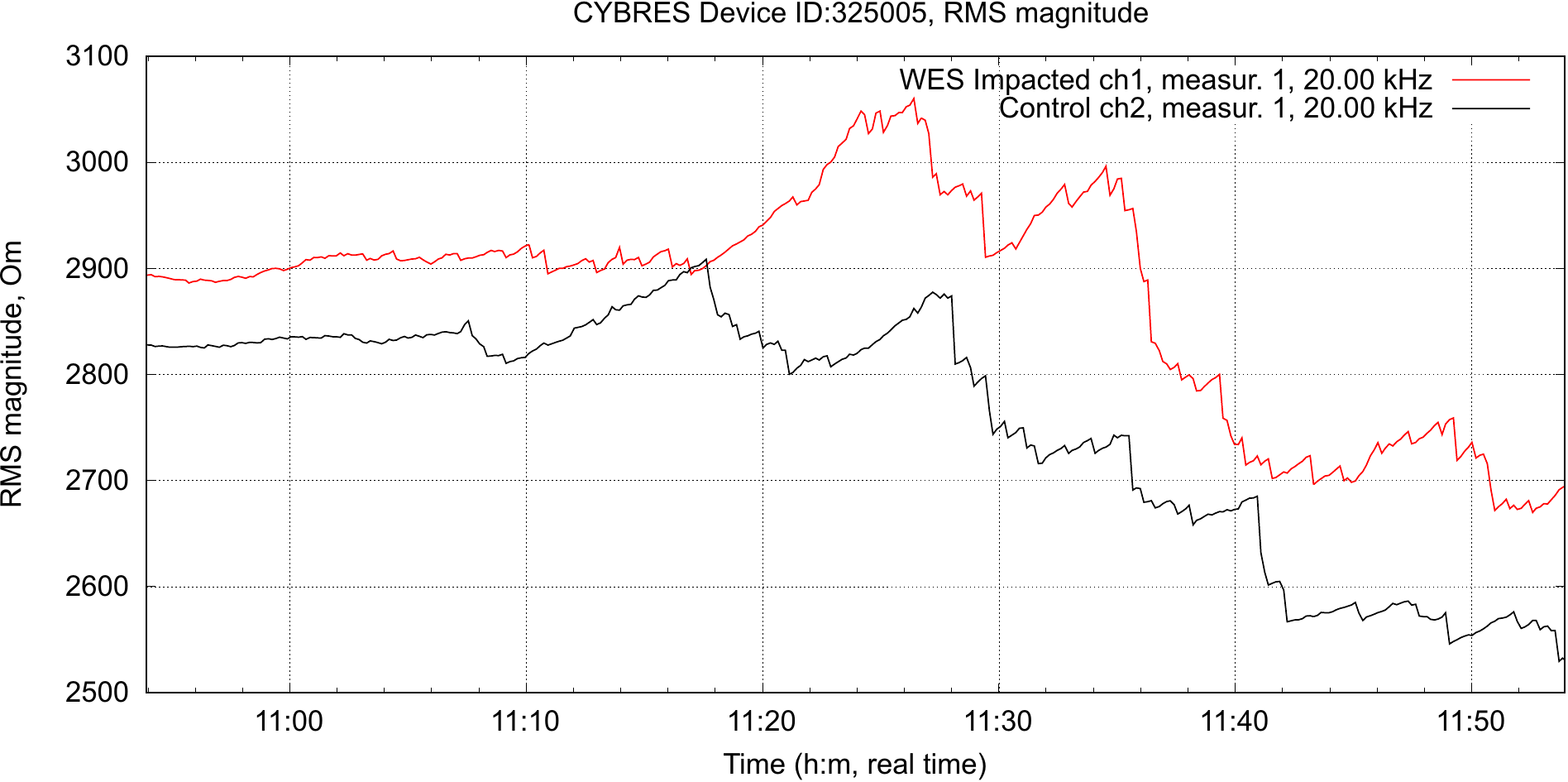}}~
\subfigure[\label{fig:MEMONdyn2}]{\includegraphics[width=.49\textwidth]{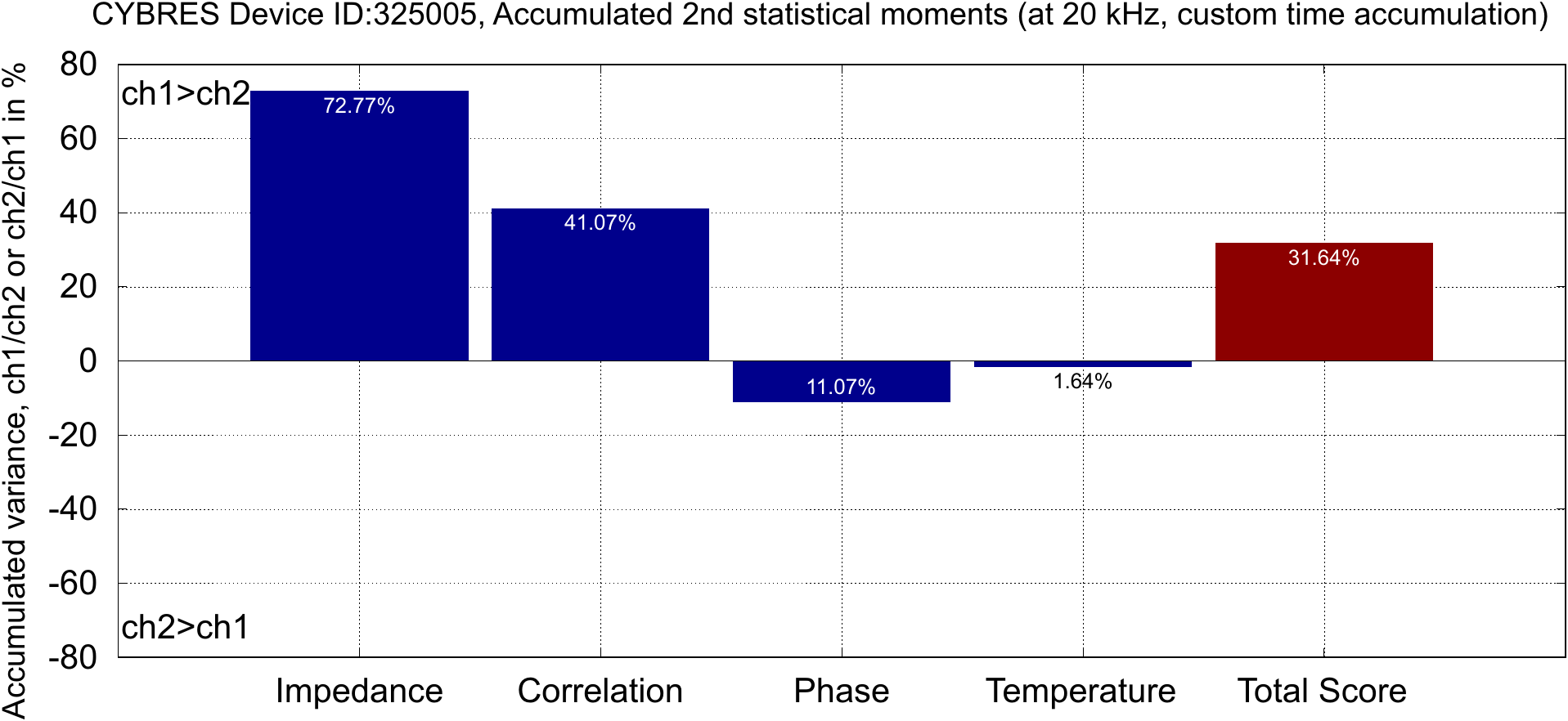}}
\subfigure[\label{fig:WESdyn21}]{\includegraphics[width=.49\textwidth]{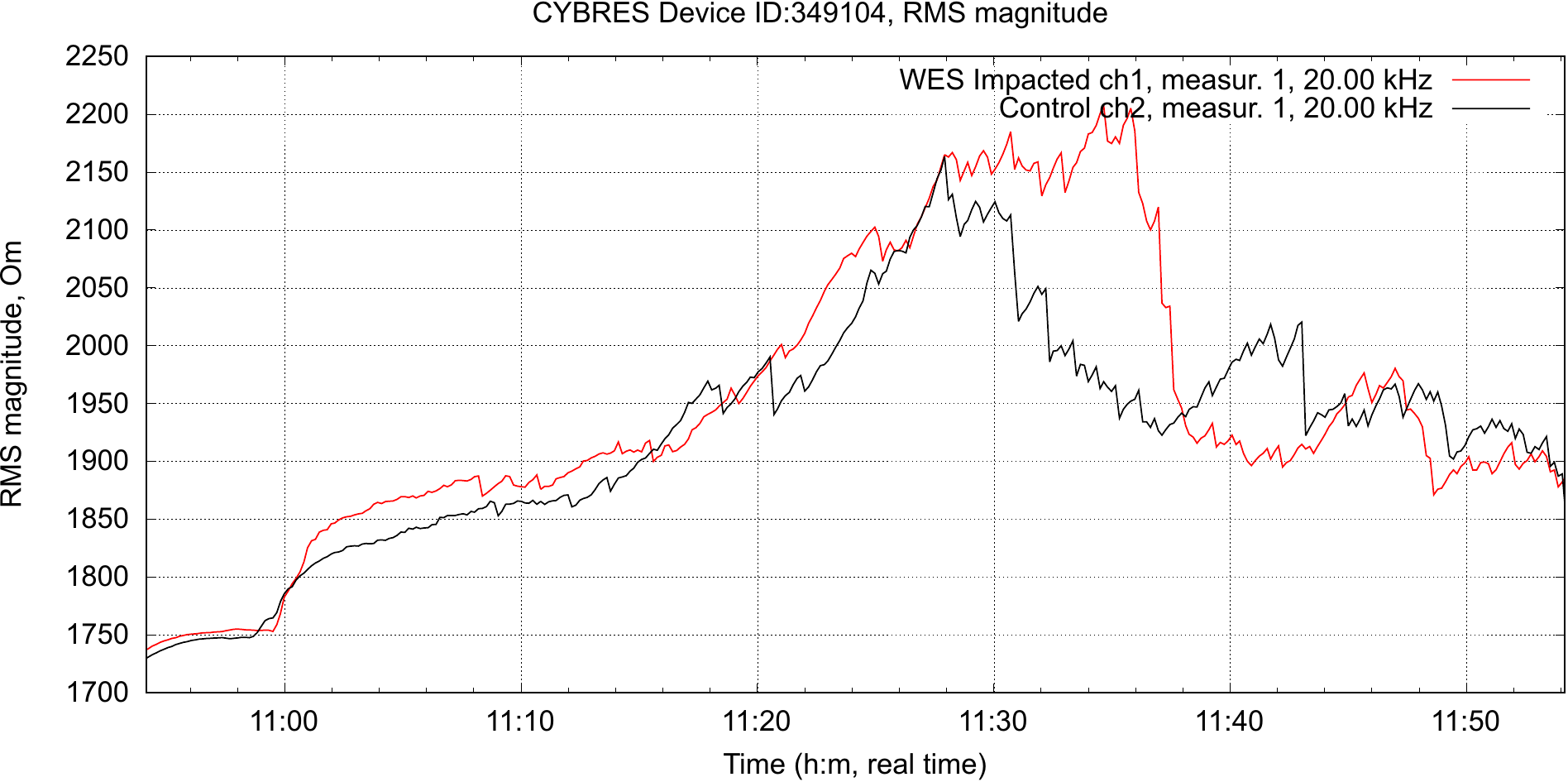}}~
\subfigure[\label{fig:WESdyn22}]{\includegraphics[width=.49\textwidth]{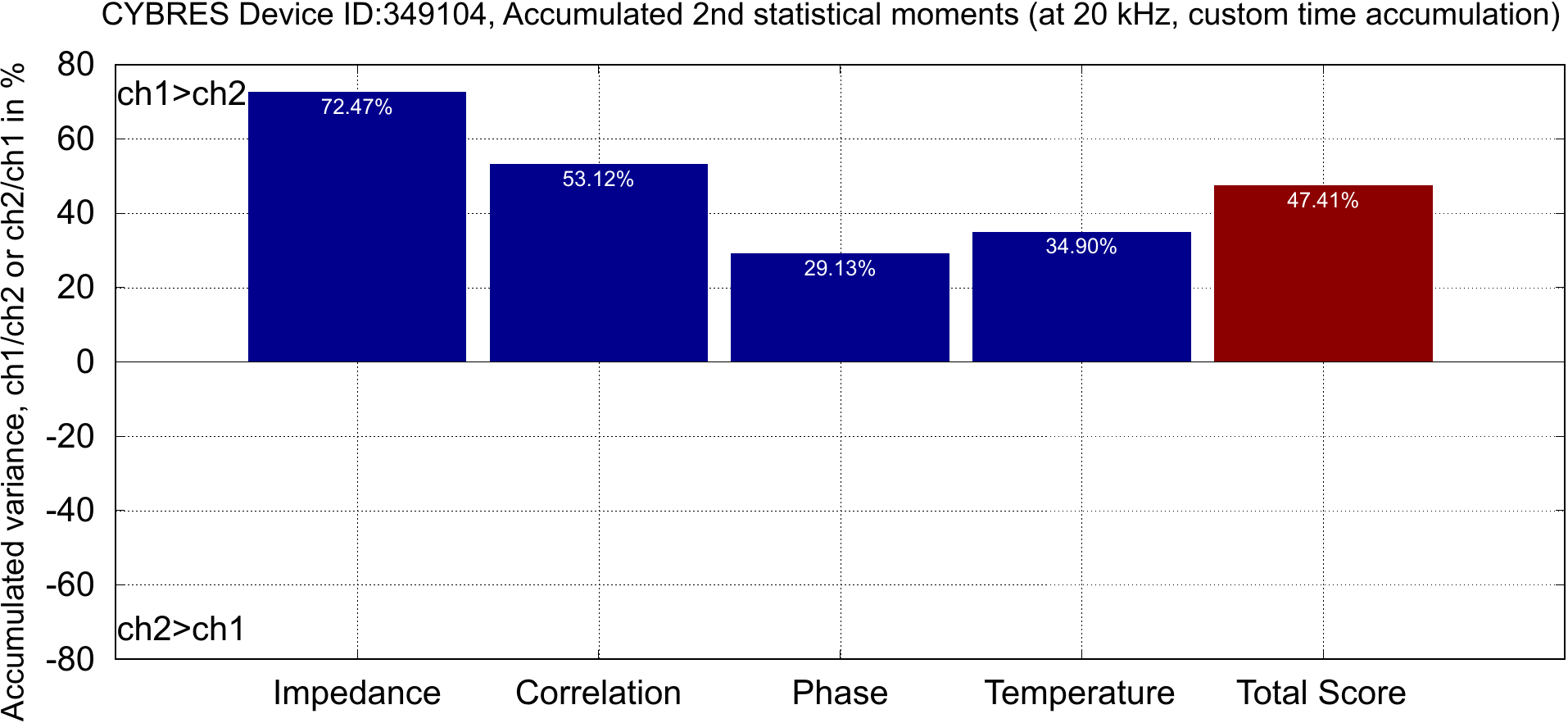}}
\subfigure[\label{fig:lowFreq1}]{\includegraphics[width=.49\textwidth]{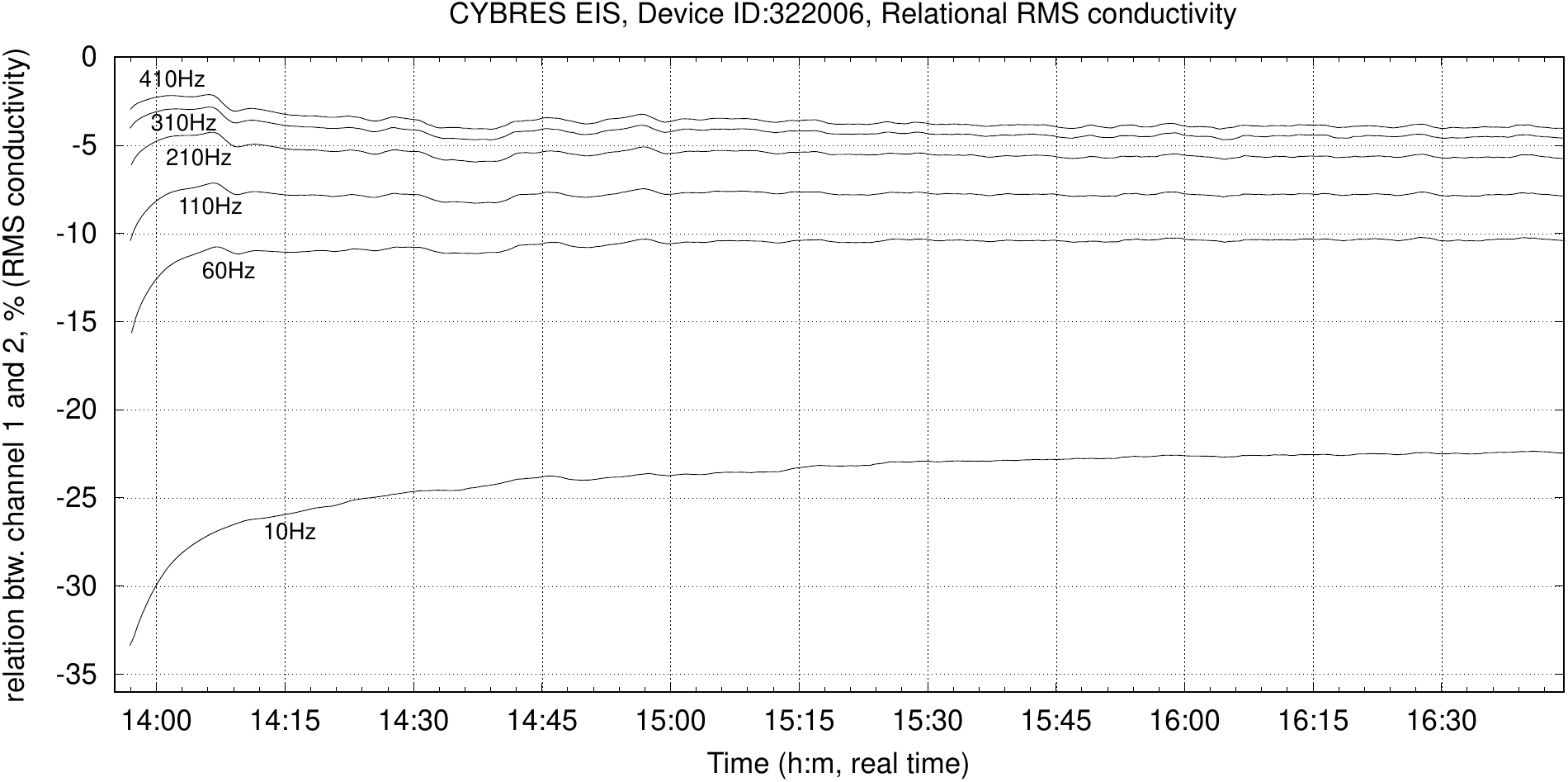}}~
\subfigure[\label{fig:lowFreq2}]{\includegraphics[width=.49\textwidth]{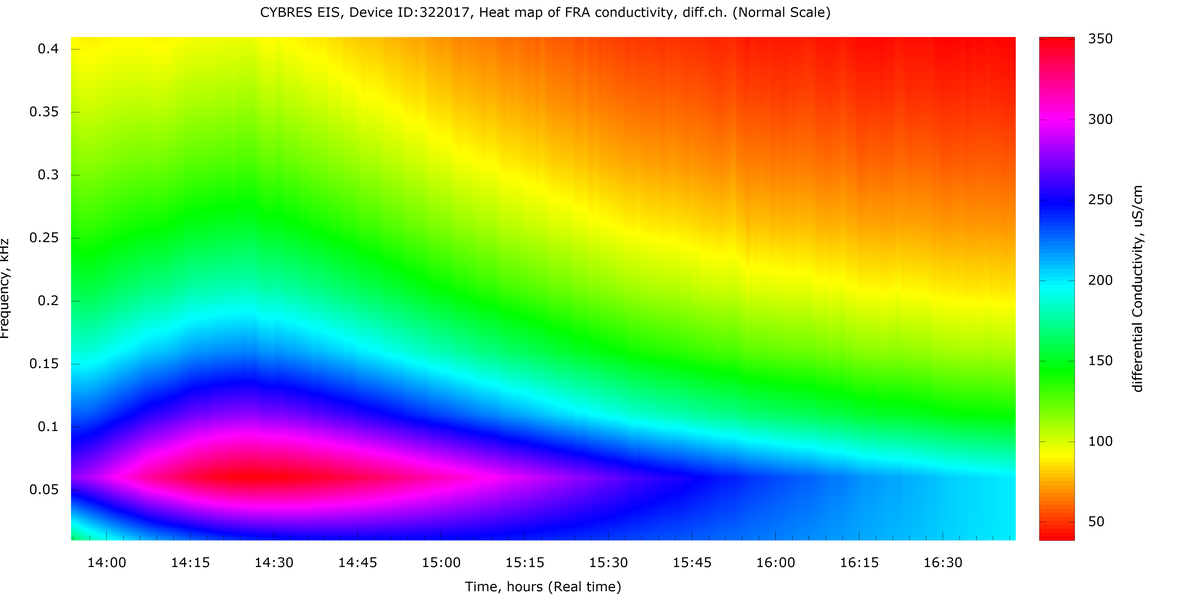}}
\caption{\small Yeast fermentation EIS dynamics at \SIlist{20}{\kilo\hertz}. \textbf{(a)}  Control measurement: RMS magnitude and \textbf{(b)} Accumulated variance; \textbf{(c)} Experimental attempt with WES impact (memonizerCOMBI) in Ch1: RMS magnitude and \textbf{(d)} Accumulated variance; \textbf{(e)} Experimental attempt with WES impact (orthogonal magnetic and electric fields) in Ch1: RMS magnitude and \textbf{(f)} Accumulated variance; \textbf{(g)} EIS dynamics at \SIlist{10;60;110;210;310;410}{\hertz} and \textbf{(h)} Time-frequency patterns appearing at low (\SI{< 0.5}{\kilo\hertz}) frequencies of EIS measurements.
}
\end{figure*}

\begin{figure}[h!]
\centering
\subfigure{\includegraphics[width=.49\textwidth]{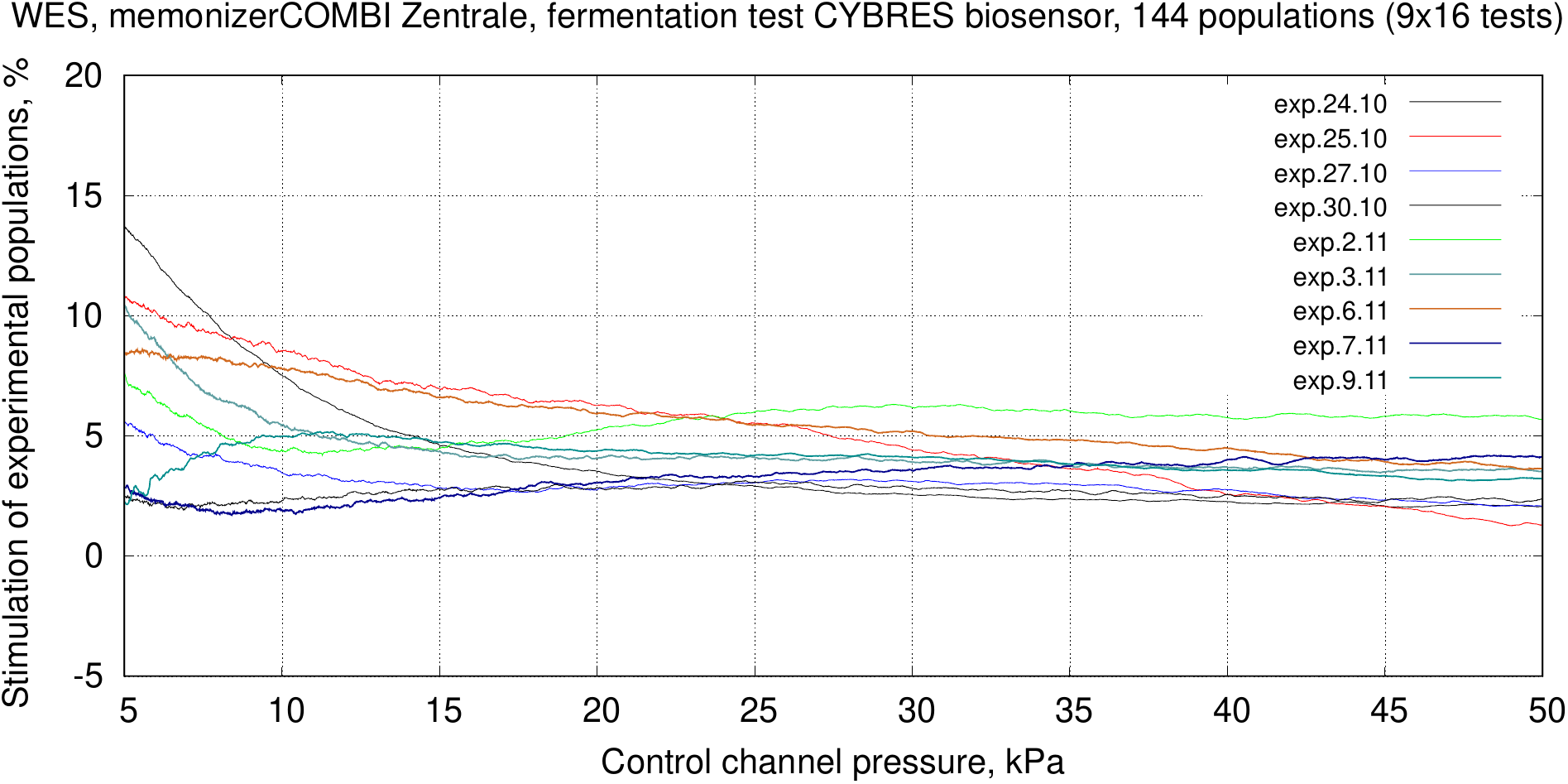}}
\caption{\small Nine repeating experiments with WES (EM emission from power lines with two spiral luminescent lamp \SI{45}{\watt} as load) and the system 'memonizerCOMBI Zentrale', see description in text. \label{fig:memon}}
\end{figure}

\subsection{Environmental stressors}

These experiments for sensing WES were related to electric and magnetic fields. In first experiments we selected \SI{50}{\hertz} power lines as a source of EM emission. The setup consisted of two such power lines placed on the table with about \SI{1}{\meter} distance, the load represented by two spiral luminescent lamp \SI{45}{\watt}, connected to each power line. The measured intensity of alternating electric field in such conditions is about \SI[per-mode = symbol]{120}{\volt\per\meter} and magnetic field \SI{650}{\nano\tesla} (measured by Spectran NF 5010). Two containers with yeast (without sugar) are placed directly on these lines, the exposition time was \SI{4}{\hour}. To create a symmetry breaking setup, e.g. to show the difference between control and experimental populations under WES impact, we placed one of the available on the market EM-protection devices\footnote{The device 'memonizerCOMBI Standard' produced by 'memon\textsuperscript{\textregistered} bionic instruments GmbH'.} on one power line about \SI{10}{\centi\meter} away from the experimental population. The null-hypothesis is that equal conditions in both lines should not produce any differences in samples. Rejecting null-hypothesis means that WES factors, produces by e.g. EM-protection devices, will be reflected in appearance of symmetry breaking. Nine measurements on different days were performed -- in total 144 populations, 8 populations were discarded, the homogeneity of these measurements is \SI{94.5}{\%}. In these attempts we observed about 4.5\% of stimulation of experimental populations at pressures \SI{> 25}{\kilo\pascal}. This symmetry breaking implies rejecting the null hypothesis and can be interpreted in several ways, e.g. towards estimating the biological effects of EM emissions, or testing the functionality of EM protection devices as it was intended in this case. Fig. \ref{fig:MEMONdyn1} shows the EIS dynamics of the same WES impacted yeast fermentation process. The difference between channels is shown in the accumulated variance bar diagram. The 'Total Score' difference is about \SI{31.5}{\%}.

Separate line of these experiments represents the vector magnetic potential, which is a three-dimensional vector field whose curl is the magnetic field (B-field). Rampl at al. described in details the phenomenon and its influence on living cells \cite{Rampl2012}. We made attempts to replicate these experiments with some modifications for orthogonal magnetic and electric fields. One of two identical 100 ml containers with sugar and 'Simeticon' solution was exposed by such system about 1 hour. Then both containers were put in the water bath to exclude any temperature difference before the yeast injection. The result of measurements is shown in Fig. \ref{fig:WESdyn21}. The 'Total Score' difference between control and impacted channels is \SI{47.41}{\%}.

\subsection{Sedimentation and double electric layer measurements}

The impedance of fermentation represents an informative parameter for the frequency range below \SI{10}{\mega\hertz} \cite{Yardley2000}. The frequencies lower than \SI{1}{\kilo\hertz} show the impedance proportionality to the biomass change in the batch. The measurements in this frequency range are largely determined by the double-layer region around electrodes and rather minor interferences with technological parameters (e.g. aeration and stirring). The comparison of different biomass measurement approaches (dry cell weight—DCW and optical density—OD) proves a good correlation of double-layer capacitance with the cell density. However, the active phase of cell division starts after 4-6 hours of the fermentation process \cite{Slouka2017} and can be excluded from the list of the influencing factors. The relatively fast impedance changes in the 1st hour of experiments are caused by the sedimentation process and environmental factors \cite{Soley2005}. Ebina et al. demonstrated that the impedance changes by ionic variation during the yeast metabolism and growth process \cite{Ebina1989}.

As shown in Fig. \ref{fig:EIS2a}, the EIS measurements at \SI{< 0.5}{\kilo\hertz} are not suitable for estimating the ionic content of fermentation dynamics. For these measurements we recommend the frequency range \SIrange[range-phrase = --]{5}{10}{\kilo\hertz} that produces a minimal measurement noise. However, the low frequency measurements can be of interest for two specific cases. Figs. \ref{fig:lowFreq1}, \ref{fig:lowFreq2} show measurements at \SIrange[range-phrase = --]{10}{410}{\hertz}, we can observe more significant changes at \SI{< 100}{\hertz} during first 10 minutes when the sedimentation takes place. Thus, this frequency range can be useful for measuring the sedimentation processes. The second specific case is related to measurements based on double electric layers with DC current in pure water. As shown e.g. in \cite{Kernbach12JSE}, this approach possesses a high sensitivity to WES. It seems that AC current at frequencies \SI{< 10}{\hertz} for colloidal fluids produces similar polarization effects and can be used for such measurements.

\section{Conclusion}
\label{sec:conclusion}

This work represents the approach based on the zymase activity of yeast \emph{Saccharomyces Cerevisiae} for estimating the biological impact of different factors: chemical contamination, the quality of water, non-chemical fluid processing, weak environmental stressors. Three different devices, see Fig. \ref{fig:Devices} -- based on 16-channel pressure sensing and 2-channel electrochemical impedance measurement with \SI{15}{\milli\litre} and \SI{100}{\milli\litre} containers -- were developed and tested for these purposes. Beside standalone applications, the EIS approach is intended as a sensor for different collective robotic systems \cite{Kernbach08online}, \cite{Kornienko_S05e} also for underwater environments \cite{Thenius16}.

We estimate the repeatability of pressure measurements in average case of about \SI{+-1}{\%} for pressures \SIrange[range-phrase = --]{20}{30}{\kilo\pascal}, lower pressures have a higher variation of results due to different factors such as unstable 'lag-''phase of fermentation, combination of aerobic and anaerobic types, initial variation of temperature. Fermentation behavior at higher pressures is inhibited by produced ethanol and also by $CO_2$ stress, due to these factors the difference between control and experimental channels decreases for pressures \SIrange[range-phrase = --]{> 30}{35}{\kilo\pascal}. The EIS dynamics demonstrates a similar behaviour, however it has a lower difference between channels due to ethanol and $CO_2$ that simultaneously increase and decrease the ionic content. We estimate the noise level of EIS dynamics at \SI{+-0.5}{\%} between channels, the values over \SI{+-0.5}{\%} boundary are assessed as significant.

The performed measurements demonstrate about \SIrange[range-phrase = --]{30}{55}{\%} of difference between control and experimental channels for tested chemical contaminations, about \SIrange[range-phrase = --]{5}{25}{\%} for tested types of water. Weak stressors have lower values -- about \SIrange[range-phrase = --]{2.5}{5}{\%} with pressure sensing. The EIS sensing provides $\num[separate-uncertainty]{18.71(783)}$\% of 'Total Score' for control attempts, the performed tests indicated between 30\%-50\% for WES and $>150\%$ for $CuSO_4$ and $C_2H_5OH$ chemical contaminations. The EIS can provide an additional insight into the fermentation process, e.g. the set point selected at the transition between phases B and C, see Fig. \ref{fig:EIS}, that can be used for identification of chemical and biological processes in these phases.
\begin{figure}[h!]
\centering
\subfigure{\includegraphics[width=.49\textwidth]{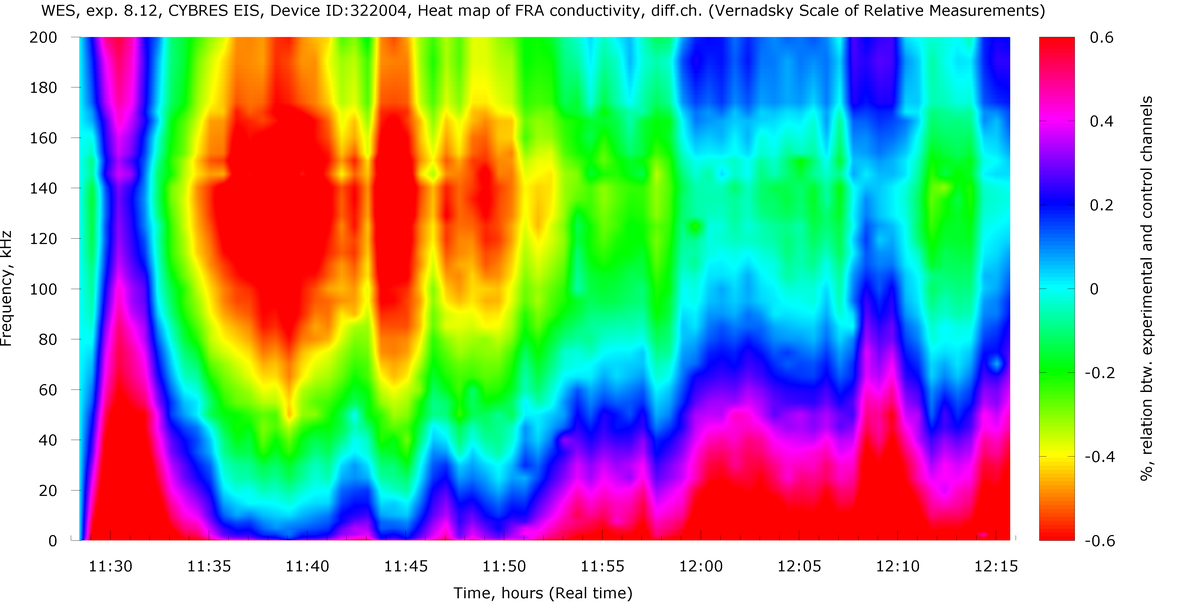}}
\caption{\small EIS dynamics from Fig. \ref{fig:EIS}, the set point is selected at the transition between phases B and C. The appearing time-frequency patterns can be used for identification of chemical and biological processes in these phases. \label{fig:differentFrequenciesEIS}}
\end{figure}

The multichannel pressure sensing system has an advantage of a fast accumulation of statistically significant results. However, the drawback of this approach consists in a larger variation of experimental and control populations that increases a random inaccuracy of measurements. These variations can be decreased by post-experimental data processing e.g. by switching off 'abnormal' channels, however this cannot be used as a regular procedure. Better approach is to perform averaging in a 'physical way', e.g. to use large EIS measurement containers (\SI{100}{\milli\litre} and more) to increase a homogeneity of yeast solution. This enables reducing the number of channels only to differential sensing and to decrease the cost and complexity of measurements. Comparing the pressure sensing and EIS we see a clear advantage of EIS (with large containers). Beside a simpler treatment of samples, the EIS enables sensing of ionic content, the sedimentation process, the yeast biomass, and effects based on double electric layers -- it provides more information for characterising environmental measurements and different weak stressors.

\section{Acknowledgement}

This work is partially supported by the FET Innovation Launchpad Project: 'E-SPECTR: Excitation Spectroscopy Sensor', grant No: 800860, EU-H2020 Projects 'subCULTron: Submarine cultures perform long-term robotic exploration of unconventional environmental niches', grant No: 640967 and 'WATCHPLANT: Smart Biohybrid Phyto-Organisms for Environmental In Situ Monitoring', grant No: 101017899 funded by European Commission.

\small
\IEEEtriggeratref{63}
%\bibliographystyle{unsrt}
%\bibliography{bibliography.tex}

\end{document}